\documentclass[aps,amsmath,preprint,amssymb,floatfixng,showpacs,
superscriptaddress,footinbib]{revtex4-2}
\usepackage[T1]{fontenc}
\usepackage{bm}
\usepackage{graphicx}
\usepackage{graphics}
\usepackage{mathtools}
\usepackage{newtxtext} 
\usepackage{amsmath}
\usepackage{hyperref}
\hypersetup{
    colorlinks,%
    citecolor=blue,%
    linkcolor=blue,%
    urlcolor=blue
}
\usepackage{tikz}\usetikzlibrary{petri}
\usepackage{soul}

\begin{document}


\title{Alloy engineering of Magnetic phases in two-dimensional Chromium Trihalides}


\author{Pedro Roberto Lopes Vieira}

\affiliation{Institute of Physics, University of São Paulo, São Paulo, São Paulo, Brazil.}

\author{Daniel D. Rivera}
\affiliation{Institute of Physics, University of São Paulo, São Paulo, São Paulo, Brazil.}

\author{Lucas Martin Farigliano}
\affiliation{Institute of Physics, University of São Paulo, São Paulo, São Paulo, Brazil}

\affiliation{Department of Theoretical and Computational Chemistry, National University of Córdoba, Córdoba, Argentina.}

\author{Fernando P. Sabino}
\affiliation{Department of Materials Engineering, University of São Paulo, São Carlos, São Paulo, Brazil.}

\author{Gustavo Martini Dalpian}
\email[]{dalpian@usp.br}
\affiliation{Institute of Physics, University of São Paulo, São Paulo, São Paulo, Brazil.}


\date{\today}

\begin{abstract}
Two-dimensional magnetic materials offer unique opportunities for exploring low-dimensional spin phenomena and hold some potential for next-generation spintronic devices due to their tunable magnetic properties and atomically thin form factors. Chromium trihalides {CrX}$_{3}$ (X = Cl, Br, I) belong to an important family of two-dimensional magnetic materials. Studying alloys in {CrX}$_{3}$ systems opens important opportunities for tailoring their electronic, magnetic, and optical properties, enabling precise control over material behavior, including its stability. In this work, we present a first-principles theoretical study, based on density functional theory (DFT), of chromium trihalides {CrX}$_{3}$ compounds and their ternary alloys.  Our results show that for the pure compounds, the ground state is intrinsically ferromagnetic (FM), with the antiferromagnetic-zigzag (AFM-Z) and paramagnetic (PM) phases being close in energy. Still, for pure compounds, we observe that the band gap variation among different magnetic phases does not exceed 0.16 eV, whereas the average magnetic moments on chromium atoms is also very similar for all magnetic phases, and increases from Cl to Br and to I compounds. For the alloys, the FM state remains the lowest-energy configuration, but the energy difference towards the AFM-Z phase decreases for compounds with lower iodine concentration. The band gaps were calculated, revealing a pronounced bowing along the compositional edge connecting CrCl$_3$ and CrI$_3$. The Curie temperatures were also determined, showing a smooth variation across compositions, consistent with the nearly linear behavior of the magnetic exchange parameters along the edges connecting the parent compounds. Based on the calculated mixing enthalpy and configurational entropy, we obtained an approximate Gibbs free energy, indicating that alloy formation becomes thermodynamically favorable at finite temperatures. This observation is important to help overcome the intrinsic experimental instability of these compounds. Our findings provide insight into the interplay between structural, electronic, and magnetic properties in {CrX}$_{3}$ alloys, offering ways to tune its properties towards targeted properties.
\end{abstract}

\keywords{Two-dimensional magnetic materials, Chromium trihalides, Density functional theory, Alloys, Spintronics.}

\maketitle

\section{Introduction \label{Introduction}}
Two-dimensional (2D) magnetic materials have garnered immense attention over the last decade due to their unique electronic, mechanical, and optical properties, which often differ significantly from those of their three-dimensional counterparts\cite{Kumari2021,Khan2020,Hossain2022,Mi2023,Elahi2022,Acosta2022}. For a long time, the existence of magnetic order in two-dimensional systems was considered unlikely due to the Mermin-Wagner theorem\cite{Mermin1966}, which predicted the absence of long-range magnetic ordering in ideal 2D systems at finite temperatures. However, the recent discovery of 2D magnetic materials, such as CrI$_{3}$\cite{Huang2017} and Cr$_{2}$Ge$_{2}$Te$_{6}$\cite{Xing2017}, has shown that spin anisotropy\cite{Ren2021}, long-range interactions\cite{Baranava2020}, and other complex correlations can stabilize magnetism in monolayers and multilayers of these systems\cite{Liu2022,Song2019,Menichetti2024,Verzhbitskiy2020,Menichetti2019,Sun2018}. In addition to promoting scientific advances, the study of 2D magnetic materials can lead to technologies that require lower energy consumption and offer greater efficiency, thereby contributing to sustainable innovation and the development of devices that minimize environmental impact.  Moreover, their compatibility with other van der Waals materials opens pathways for the fabrication of heterostructures with novel magnetic and electronic properties\cite{Han2024,Paolucci2024,Cheng2022,Ghazaryan2018,Wu2021,Mukherjee2020}.

Among a large variety of 2D materials, chromium trihalides (CrX$_{3}$, where X = Cl, Br, I) are a family of layered magnetic materials that have garnered significant attention due to their intriguing magnetic properties and potential applications in spintronics\cite{Soriano2020,Zhang2022,Acharya2021,Soriano2022,Xu2020,Ghader2022,Esteras2023,Basak2023}, due to their tunable magnetic properties and potential for hosting exotic magnetic states such as spin spirals and topological magnetic phases\cite{Heinrich2021,Baidya2018}. 

These compounds crystallize in a layered structure where the chromium atoms are surrounded by halogen atoms forming edge-sharing octahedra, and where different layers interact through van der Waals interactions\cite{Huang2017,Yao2023,Kazim2020}. This structural arrangement allows for easy exfoliation into monolayers, making them ideal candidates for studying two-dimensional magnetism\cite{Xu2021,Wang2021,Zhang2015,Hasan2023,Bacaksiz2021}. The magnetic properties of chromium trihalides arise from the Cr$^{3+}$ ions, which have a partially filled 3d shell resulting in a magnetic moment of $\sim$ 3 $\mu_{B}$ per ion. These materials display a variety of magnetic phases depending on the halide composition and the temperature\cite{Chen2019}. Experimentally, for its two-dimensional form, CrCl$_3$, CrBr$_3$, and CrI$_3$ exhibit ferromagnetic behavior with Curie temperatures (T$_{c}$) of approximately 15 K\cite{BedoyaPinto2021,crcl3arxiv}, 37 K \cite{Tsubokawa1960}, and 61 K\cite{McGuire2015,Hansen1959}, respectively.

The magnetic and electronic properties of this class of materials can become even more interesting with the creation of alloys by modifying the sites that contain the halides, giving rise to compounds with stoichiometry Cr(Cl$_{x}$Br$_{y}$I$_{1-x-y}$)$_{3}$. Studying these alloys can provide insights into how the modulation of chemical composition can alter the band structure, changing the material's energy gap. Different halogens have varying effects on the electronic density of states, which can influence conduction and electrical insulation properties, as well as potentially affect the system's structural stability by altering atomic bonds, crystal lattice geometry, and thermodynamic properties.\cite{Sabino2025} It has already been reported in the literature that alloys similar to those in our study can significantly modify properties such as the collinear exchange constants of the nearest neighbors (and the next nearest neighbors), the onsite anisotropy strength D, the Curie temperature, and the easy magnetic axis\cite{Chen2021}. Therefore, chromium trihalides and its alloys are a versatile platform for exploring low-dimensional magnetism.

This kind of strategy can also help to make these two-dimensional compounds more stable, similar to what happens in high entropy alloys\cite{George2019}. Entropic contributions owing to the diversity of positional configurations in these alloys contribute to decrease the Gibbs free energy.   CrX$_{3}$ are knows to have limited stability\cite{Zhang2022_2,Mastrippolito2021,Shcherbakov2018}, and alloying can be a good strategy to overcome these issues. 

Here, we will use quantum mechanics methods to study the properties of CrX$_{3}$ and we extended our study to ternary alloys. We studied how the magnetic properties of these compounds change with the different magnetic phases and the generation of different alloys in the halogen atoms. We have observed that for all the materials the ground state is intrinsically ferromagnetic (FM), with the antiferromagnetic-zigzag (AFM-Z) and paramagnetic (PM) phases being close in energy. For the pure compounds the band gap variation among different magnetic phases does not exceed 0.16 eV, while the average magnetization magnitude on chromium atoms remains within 0.05 $\mu_{\mathrm{B}}$ between different magnetic phases. For the alloy systems, heatmaps reveal that compositions with reduced iodine concentrations exhibit smaller energy differences between the FM ground state and the AFM-Z phase, remaining below 6 meV/f.u. 
Furthermore, bowing behavior is observed in both the band gap and the mixing enthalpy across the ternary alloys formed by CrCl$_3$, CrBr$_3$, and CrI$_3$. Although non-linear deviations are present throughout the compositional space, they are particularly pronounced along the CrCl$_3$–CrI$_3$ direction. These trends are clearly visible in the heatmaps, highlighting the interplay between halogen substitution and the resulting structural and electronic responses. Our findings provide insight into the interplay between structural, electronic, and magnetic properties in {CrX}$_{3}$ alloys, with potential implications for spintronic applications. 

\section{Computational details \label{Computational details}}

To investigate the structural, electronic, and magnetic properties of CrX$_{3}$ compounds and their ternary alloys, we performed first-principles calculations based on density functional theory (DFT) \cite{Hohenberg1964,Kohn1965} as implemented in the Vienna Ab initio Simulation Package (VASP)\cite{Kresse1993}, version 6.4.2. To solve the Kohn-Sham equation, the projected augmented wave (PAW) method\cite{Blchl1994,Kresse1999} was employed, and the valence electrons in each atom are shown in parentheses: Cr (3d$^{5}$ 4s$^{1}$), Cl (3s$^{2}$ 3p$^{5}$), Br (4s$^{2}$ 4p$^{5}$) and I (5s$^{2}$ 5p$^{5}$). We used the non-local generalized gradient approximation (Meta-GGA) r$^{2}$SCAN\cite{Furness2020} for the exchange-correlation energy functional. The equilibrium 2D lattice parameters were obtained by stress-tensor minimized on the plan of the layered CrX$_{3}$ (X = Cl, Br and I) and forces minimization in all the directions, with a cutoff energy of 400 eV. A vacuum thickness of 20 Å separates the periodic images of the layers, avoiding interactions between them. We investigate the AFM-Néel (AFM-N), AFM-stripe (AFM-S), AFM-zigzag (AFM-Z), Ferromagnetic (FM), and Paramagnetic (PM) phases for the pure compound. It is worth highlighting that the PM phase has not been investigated in this way for these compounds before. Some of these magnetic phases have already been reported in the literature for different materials\cite{An2022,Kan2013,Wang2011,Chittari2020,Chen2020}. All magnetic configurations can be found in the Supplementary Material (SPM). For the FM phase, structural relaxations were performed using a k-point mesh of $6 \times 6 \times 1$ for CrCl$_3$, $5 \times 5 \times 1$ for CrBr$_3$, and $5 \times 5 \times 1$ for CrI$_3$. For the projected density of states (PDOS) calculations, the corresponding meshes were $10 \times 10 \times 2$, $9 \times 9 \times 2$, and $8 \times 8 \times 2$ for CrCl$_3$, CrBr$_3$, and CrI$_3$, respectively. For the remaining systems, the k-point meshes were scaled according to the cell sizes to ensure a consistent k-point density across all calculations.
\begin{figure}[htb]
       \centering
      \includegraphics[width=155mm]{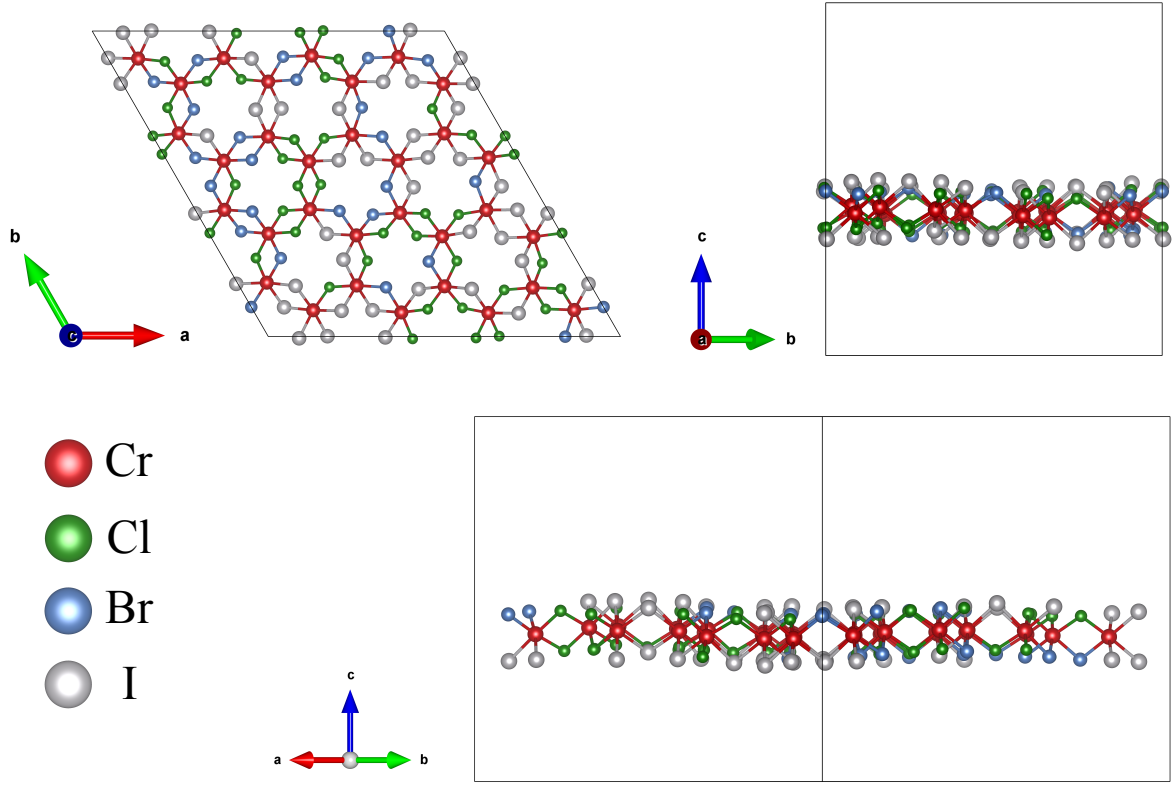}
         \caption{Alloy Cr(Cl$_{0.375}$Br$_{0.25}$I$_{0.375}$)$_3$, corresponding to the composition (25.0, 37.5, 37.5). Red, green, blue, and gray spheres represent Cr, Cl, Br, and I atoms, respectively. The alloy supercell corresponds to a 4$\times$4 expansion of the parent compounds and contains 128 atoms. The figure illustrates the random distribution of halogen atoms generated using the SQS method. }
        \label{alloy}
\end{figure}

To systematically construct all alloy configurations Cr(Cl$_{x}$Br$_{y}$I$_{1-x-y}$)$_3$, the Vegard's law was applied only on the 2D layered plan, keeping the vacuum thickness of 20 Å. According to this empirical rule, the lattice parameter of a solid solution varies proportionally to the atomic fractions of its components, assuming ideal mixing without significant atomic distortions. In this way, the lattice parameters of the alloys for each composition and magnetic phase were obtained from the relaxed lattice parameters of the pure compounds using Vegard’s law.  The halogen atoms were distributed according to the Special Quasi-Random Structures (SQS)\cite{Zunger1990,Gao2016} method implemented in Alloy Theoretic Automated Toolkit (ATAT)\cite{vandeWalle2002} software. To simulate the PM phase of all systems, we also used the SQS method applied on spins of Cr atom, as done in previous works of other types of paramagnetic materials\cite{deOliveira2020}. To calculate the T$\mathrm{_{C}}$ and the magnetic exchange parameters of the alloys, we employed the 3NN-MF method\cite{Xue2022}, which is grounded in mean-field theory and Monte Carlo (MC) studies. An explanation of this method is briefly provided in the Supplementary Material.

The halide composition was systematically varied according to these predefined concentrations, enabling a comprehensive exploration of the structural and electronic properties of mixed Cr-based halides and resulting in a total of 42 distinct alloy compositions. For example, two generated alloys were Cr(Cl$_{0.125}$Br$_{0.375}$I$_{0.50}$)$_{3}$ and Cr(Cl$_{0.375}$Br$_{0.50}$I$_{0.125}$)$_{3}$, which correspond to the (12.5, 50.0, 37.5) concentrations. Figure \ref{alloy} shows a representative example of these alloys. For each composition, we performed first-principles calculations for all magnetic phases, relaxing only the atomic positions.

\section{Results and discussion \label{Results and discussion}}

In this section, we present a comprehensive analysis of the structural and electronic properties of chromium trihalides and their corresponding alloys. First, we focus on the pristine compounds, providing a detailed comparison of their lattice parameters, electronic structures and magnetic moments. Subsequently, we explore the effects of halogen substitution on the physical properties of the {CrX}$_{3}$ alloys, with particular emphasis on how the mixture of halogens influences the energy differences between the ground state and the closest magnetic phases in energy, the band gap behavior, Curie temperatures and thermodynamic stability.

\subsection{Pure compounds: CrCl$_3$, CrBr$_3$, CrI$_3$}

Properties such as the relaxed lattice parameter, energy gap, energy difference relative to the FM phase, magnetic moments on chromium atoms and total magnetization for the pure compounds can be seen in Table I.

\begin{table*}[htbp]
    \centering
    \renewcommand{\arraystretch}{0.8} 
    \setlength{\tabcolsep}{5.7pt}
    \caption{Values obtained for the   relaxed lattice parameter $a_{0}$, energy gap E$_{g}$, energy difference with respect to the ferromagnetic phase $\Delta E_{FM}$, the arithmetic mean of the modulus of the magnetic moments $\langle | \mu_{\mathrm{m}} | \rangle$ on chromium atoms, and total magnetization for CrCl$_3$, CrBr$_3$, and CrI$_3$. The term f.u. stands for formula unit, which refers to half of the unit cell. Experimental values are shown in parenthesis\cite{Chen2019,Huang2017}.}
    \begin{tabular}{c c c c c c c}
        \hline\hline
        \textbf{System} & \textbf{Magnetic Phase} & $a_{0}$ (\AA) & $E_{g}$ (eV) & $\Delta E_{FM}/f.u.$ (meV) & $\mathbf{\langle | \mu_{\mathrm{m}} | \rangle}$ & $M_{total}/f.u.$ \\
        \hline
        {CrCl$_{3}$} 
        & \textbf{FM}     & 6.00 & 2.52 & 0.00  & 2.96 & 3.0 \\
        & \textbf{AFM-N} & 5.98 & 2.69 & 11.26 & 2.91 & 0.0 \\
        & \textbf{AFM-S} & 5.99 & 2.55 & 8.90  & 2.93 & 0.0 \\
        & \textbf{AFM-Z} & 5.99 & 2.68 & 4.67  & 2.94 & 0.0 \\
        & \textbf{PM}    & 5.99 & 2.60 & 6.27  & 2.94 & 0.0 \\
        \hline
        {CrBr$_{3}$}  
        & \textbf{FM}    & 6.37 (6.30) & 2.14 & 0.00  & 3.05 & 3.0 \\
        & \textbf{AFM-N} & 6.37 & 2.26 & 15.10 & 3.00 & 0.0 \\
        & \textbf{AFM-S} & 6.37 & 2.13 & 12.82 & 3.01 & 0.0 \\
        & \textbf{AFM-Z} & 6.37 & 2.26 & 6.88  & 3.02 & 0.0 \\
        & \textbf{PM}    & 6.37 & 2.20 & 8.97  & 3.02 & 0.0 \\
        \hline
        {CrI$_{3}$}  
        & \textbf{FM}    & 6.97 (6.87) & 1.46 & 0.00  & 3.20 & 3.0 \\
        & \textbf{AFM-N} & 6.97 & 1.59 & 19.37 & 3.15 & 0.0 \\
        & \textbf{AFM-S} & 6.97 & 1.49 & 18.49 & 3.16 & 0.0 \\
        & \textbf{AFM-Z} & 6.97 & 1.52 & 10.91 & 3.17 & 0.0 \\
        & \textbf{PM}    & 6.96 & 1.53 & 13.01 & 3.17 & 0.0 \\
        \hline\hline 
    \end{tabular}
    \label{tab:trihalides}
\end{table*}

Our results provide a FM ground state for all three materials, in agreement with the literature \cite{BedoyaPinto2021,Tsubokawa1960,McGuire2015,Hansen1959}, with the AFM-Z phase being the closest in energy, followed by the PM phase. It is worth mentioning that the energy difference between the FM and AFM-Z phases in the pure compounds is larger for CrI$_{3}$. As it is observed in Table I, the lattice parameter did not change significantly for the different magnetic configurations in the same chromium trihalides. The maximum variation was 0.02 and 0.01 \AA\ for CrCl$_{3}$ and CrI$_{3}$, respectively, while for CrBr$_{3}$ it remained constant within our precision criterion (up to the second decimal place). In the fully relaxed structures of the pure CrX$_{3}$ compounds, the Cr–Cr separations and Cr–X bond lengths systematically increase with halogen size. Specifically, the nearest‐neighbor Cr-Cr distance expands from 3.46 \AA\ in CrCl$_{3}$ to 3.68 \AA\ in CrBr$_{3}$ and 4.02 \AA\ in CrI$_{3}$. Concurrently, the Cr-X bond length increases from 2.34 \AA\ (Cr-Cl) to 2.51 \AA\ (Cr-Br) and 2.73 \AA\ (Cr-I). Despite these variations in bond distances, the Cr-X-Cr bond angles remain relatively constant, measuring 94.96° in CrCl$_{3}$, 94.31° in CrBr$_{3}$, and 94.59° in CrI$_{3}$. These results reflect the influence of halogen ionic radius on the lattice geometry.

As shown in Table I, the largest energy difference between the FM ground state and the PM phase is only 13.01 meV per formula unit. This small energy difference is consistent with the low Curie temperatures reported for these monolayers\cite{BedoyaPinto2021,crcl3arxiv,Tsubokawa1960,McGuire2015,Hansen1959}, since the Curie temperature 
T$_{c}$ is, to first approximation, proportional to a weighted sum of the exchange parameters (T$_{c}$ $\varpropto \sum _{i} z_{i} J_{i}$ where $z_{i}$ denotes the coordination number of the i-th neighbor shell and $J_{i}$
 the corresponding exchange interaction\cite{Xue2022})
At room temperature, the thermal expansion of Cr-halogen bonds reduces $\Delta_0$—the crystal field splitting between the Cr $t_{2g}$ and $e_g$ orbitals—thereby decreasing the $t_{2g}$ -  $e_g$ energy separation. This favors electron occupation of $t_{2g}$ orbitals with opposite spins, promoting antiparallel alignment and leading to the PM phase under certain conditions. However, crystal field effects should be regarded as a secondary contribution, while the dominant mechanism governing spin occupation arises from electron-electron (Coulomb) interactions.

\begin{figure}[htb]
    \centering
    \begin{minipage}[b]{0.5\textwidth}
        \centering
        \label{fig:bands_li}%
        {\includegraphics[width=82mm]{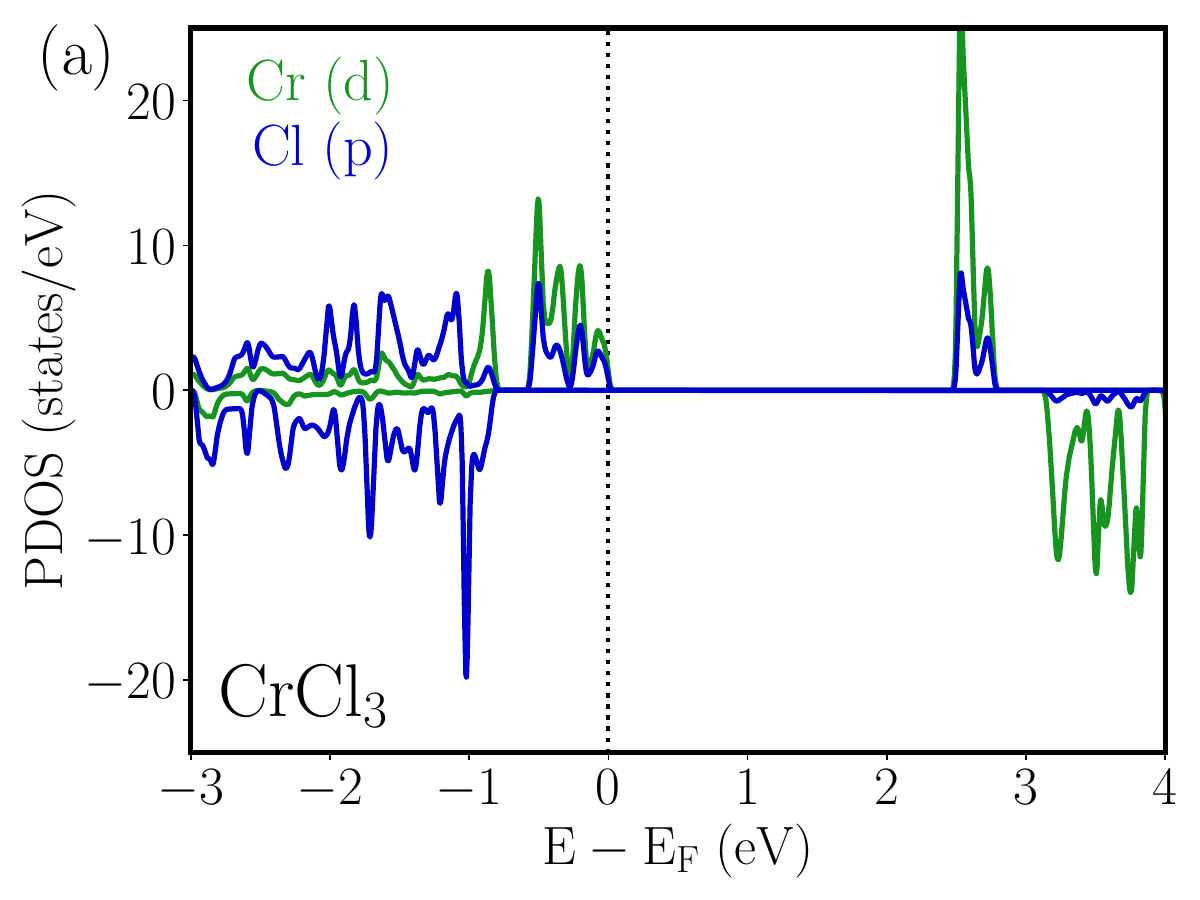}}
    \end{minipage}\hfill
    \begin{minipage}[b]{0.5\textwidth}
        \centering
        \label{fig:bands_k}%
        {\includegraphics[width=82mm]{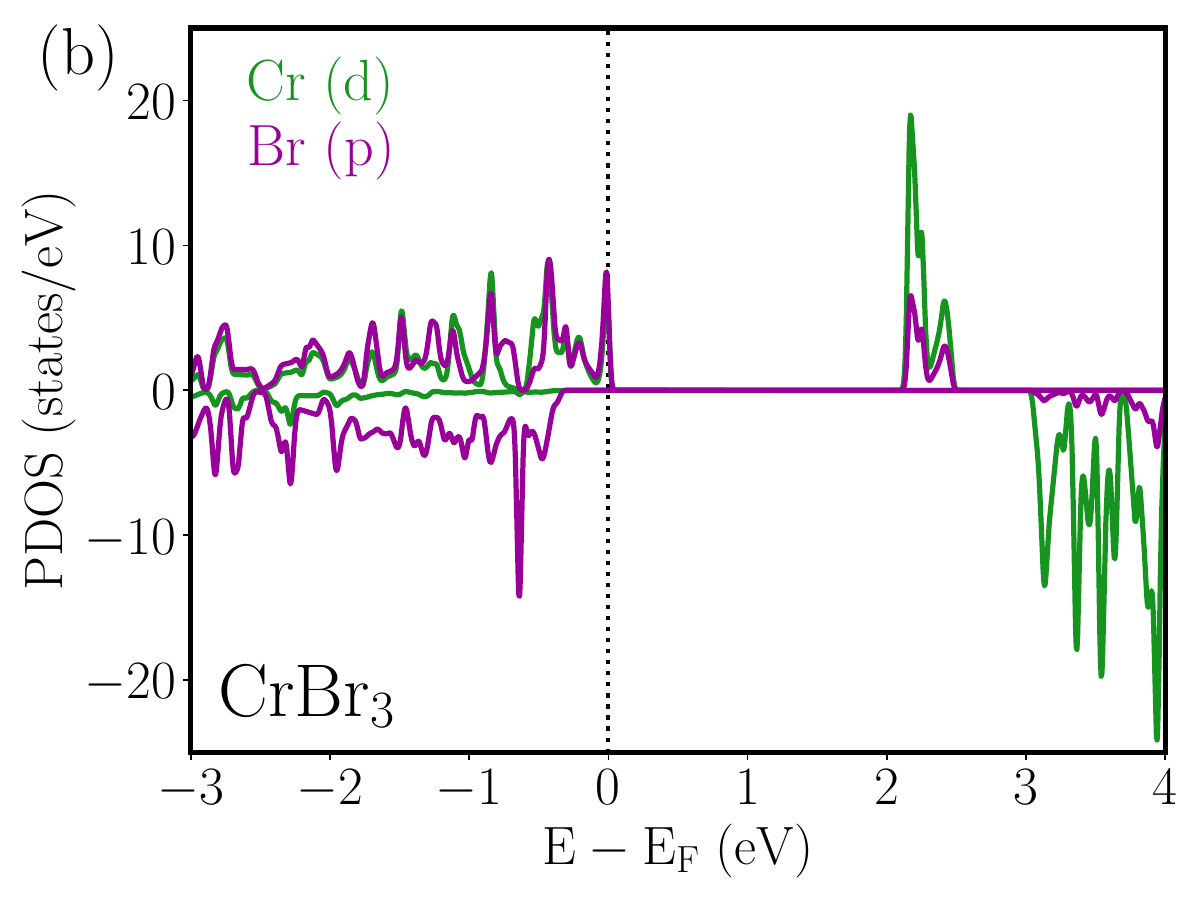}}
    \end{minipage}

    \vspace{0.4cm}

    \begin{minipage}[b]{0.6\textwidth}
        \centering
        \label{fig:bands_na}%
        {\includegraphics[width=82mm]{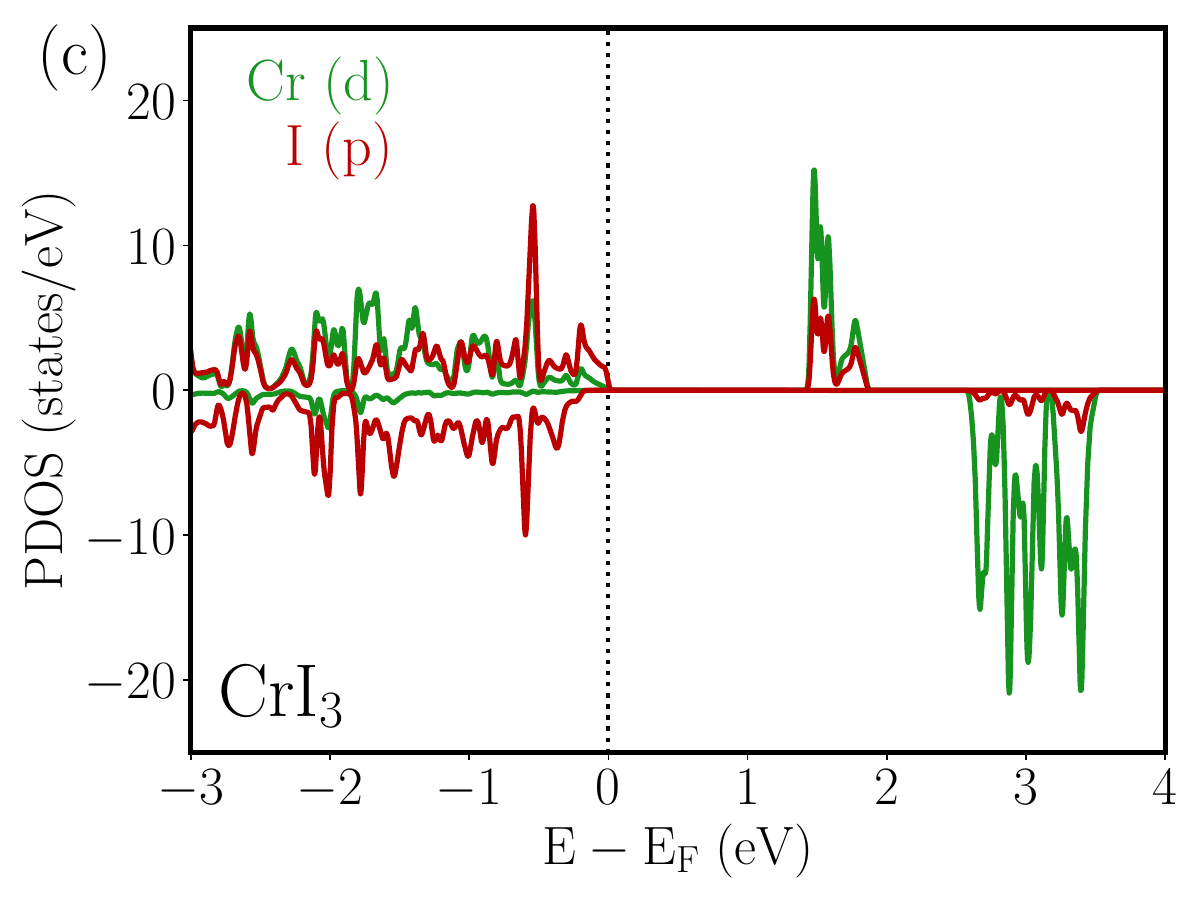}}
    \end{minipage}

    \caption{Projected density of states (PDOS) for the ferromagnetic ground state of (a) CrCl$_3$, (b) CrBr$_3$, and (c) CrI$_3$. The green curves correspond to the Cr $d$ orbitals, while the blue, purple, and red curves represent the $p$ orbitals of Cl, Br, and I, respectively. The energy zero is set at the highest occupied state.}
    \label{all_dos}
\end{figure}

Table I also shows that the average magnetic moments on Cr atoms vary with both the magnetic phase and the specific CrX$_3$ compound. This behavior arises from the hybridization between Cr d orbitals and halogen p orbitals, which redistributes electronic density and modifies the magnetic moments. Having established the structural and magnetic properties of the pure CrX$_3$ compounds, we now turn to their electronic behavior. In particular, the interplay between Cr d orbitals and halogen p orbitals, as highlighted by the variation in energy gaps, motivates a detailed analysis of the density of states. The projected density of states (PDOS) illustrates the contributions of Cr d and halogen p orbitals to the valence and conduction bands, rationalizing the trends in energy gaps and their dependence on halogen size. The corresponding projected density of states (PDOS) for the FM ground state of each compound is shown in Figure~\ref{all_dos}.

\begin{figure}[htb]
       \centering
      \includegraphics[width=140mm]{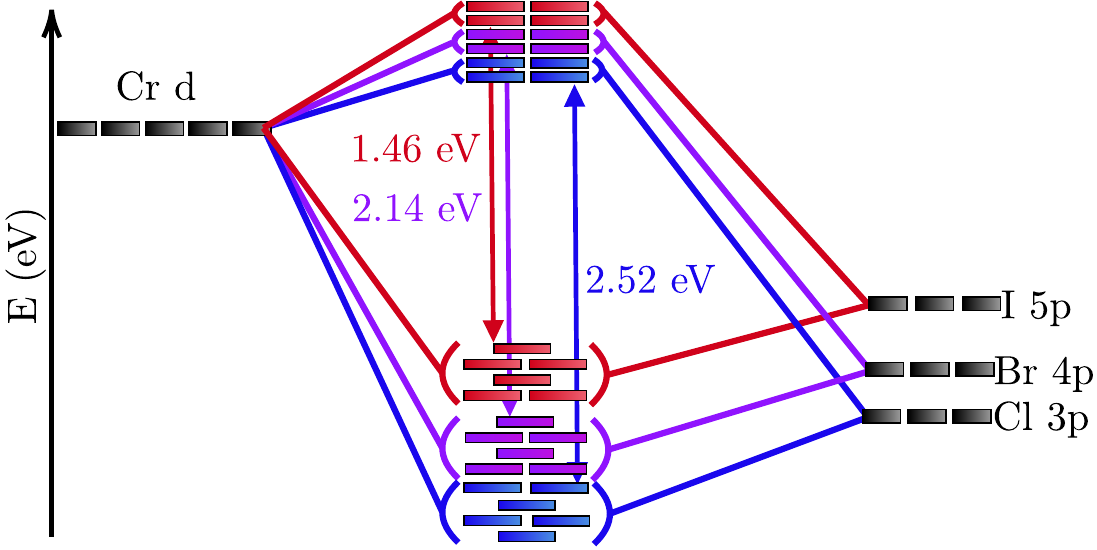}
         \caption{Molecular Orbital model illustrating the energy level alignment in CrX$_{3}$ compounds. Ligand and anti-bonding orbitals are shown in blue, purple, and red for CrCl$_{3}$, CrBr$_{3}$, and CrI$_{3}$, respectively. The arrows in the corresponding colors highlight the band gap of each material.}
        \label{orbs}
\end{figure}

Figure \ref{all_dos} shows the variation in the character of the valence band maximum (VBM) across the different halides: in CrCl$_{3}$, the Cr d orbitals dominate the VBM; for CrBr$_{3}$, there is significant hybridization between Br p and Cr d states; and in CrI$_{3}$, the I p orbitals overwhelmingly contribute to the VBM. Meanwhile, at the conduction band minimum (CBM), the Cr d orbitals remain the primary contributors across all three materials. To understand the different trends observed in the CrX$_{3}$ compounds, as shown in Table I and Figure \ref{all_dos}, a molecular orbital model is proposed. Figure \ref{orbs}, illustrates the proposed model. To build the model, the ionization energies of the neutral atoms are used as a reference. The 3p orbitals of Cl and the 3d orbitals of Cr have a large energy difference, whereas the 4p orbitals of Br and the 5p orbitals of I have intermediate energy levels and are closer in energy to the 3d orbitals of Cr, leading to strong hybridization between the orbitals at the VBM in the cases of CrBr$_{3}$ and CrI$_{3}$, as shown in Figure \ref{all_dos}.

According to crystal field theory, in an ideal octahedral environment, the Cr d orbitals split into the higher-energy bidegenerate $e_{g}$($d_{x^{2}-y^{2}}$ and $d_{z^{2}}$) and lower-energy tridegenerate $t_{2g}$ ($d_{xy}, d_{yz}$ and $d_{zx}$) orbitals. As observed in Figure \ref{orbs}, there are continuous twofold degeneracies involving hybridized $e_{g}$ and $t_{2g}$ orbitals both for VBM (a little less in the case of CrBr$_{3}$) and for CBM. This behavior arises due to distortions in the CrX$_{6}$ octahedra, which deviate from ideal symmetry. In the relaxed structures there are variations in the Cr–X bond lengths that lead to a lowering of the octahedral symmetry, causing a mixing of $e_{g}$ and $t_{2g}$ states and providing the arrangement of energy levels observed in Figure \ref{orbs}. These distortions break the strict separation $\Delta_{0}$ between the two sets of d orbitals and result in partially hybridized molecular orbitals with contributions from both symmetry types, giving rise to near-degenerate or degenerate energy levels. 

The variation in the band gaps of the CrX$_3$ compounds (Table I and Figure \ref{all_dos}) can be understood in terms of the interplay between energetic alignment and p-d hybridization \cite{Wei1998}. In general, the p-d coupling is proportional to the square of the hopping matrix element ($\varpropto |\langle X,p|\Delta V|A,d\rangle|^{2}$) and inversely proportional to the energy difference between the unperturbed Cr d and halogen p states ($\varpropto |\epsilon _{p} -\epsilon _{d} |$). Along the series Cl $\rightarrow$ Br $\rightarrow$ I, the ionization energies of neutral atoms decrease (Cl: -12.96 eV, Br: -11.81 eV, I: -10.45 eV)\cite{NIST_ASD}, indicating that the halogen p levels shift closer in energy to the Cr d states, as can be seen in Figure \ref{orbs}. This reduced energy separation tends to enhance p-d mixing and push the valence-band maximum upward. However, this effect competes with structural factors, such as increasing bond lengths and modified orbital overlap, which act to reduce the effective hybridization strength and decrease the bonding-antibonding splitting. As a result, the band gap reduction from CrCl$_3$ to CrI$_3$ arises from a balance between these competing effects, with the weakening of the bonding–antibonding splitting playing a dominant role. Consequently, CrCl$_3$ exhibits the widest gap, CrBr$_3$ an intermediate value, and CrI$_3$ the smallest gap.

\subsection{Halogen mixing alloys: Cr(Cl$_{x}$Br$_{y}$I$_{1-x-y}$)$_3$}

As previously discussed, alloying provides an effective route to tune material properties. Here, we focus on the ternary Cr(Cl$_x$Br$_y$I$_{1-x-y}$)$_3$ alloys to explore how composition affects magnetic stability, band gap, and mixing energetics. For the alloys, all magnetic phases are investigated, following the trends observed for the parent compounds. The calculations indicate that the ground state remains FM, with the AFM-Z phase being the closest in energy, followed by the PM phase. To visualize how the energy differences among magnetic phases vary across different alloy compositions, heatmaps were created. Each heatmap contains 45 points, where each point represents a distinct composition in a given magnetic phase or the value of a specific property for that composition. Each heatmap was generated using interpolated values between the discrete data points.

\begin{figure}[h!]
    \centering

    \begin{minipage}[b]{0.5\textwidth}
        \centering
        \includegraphics[width=82mm]{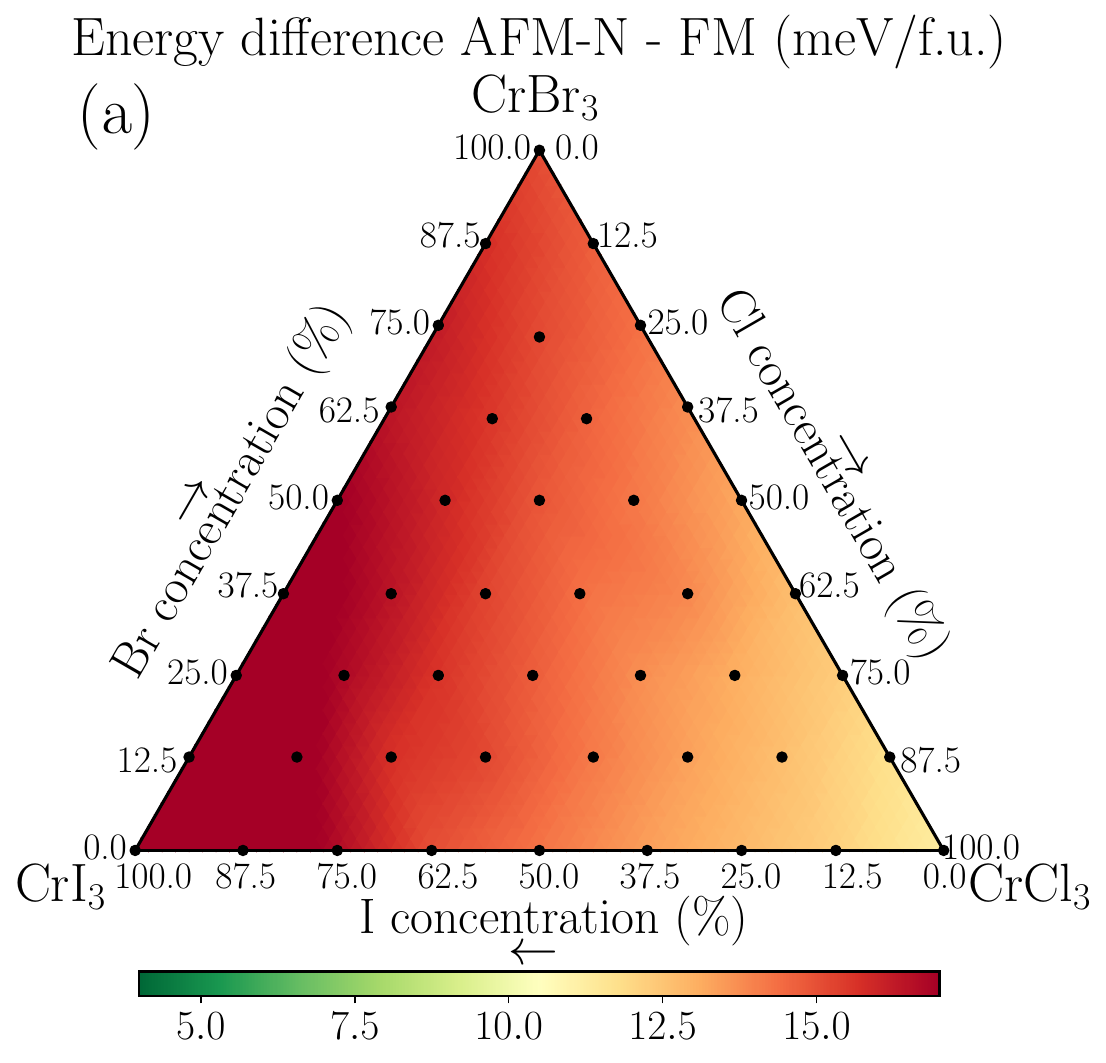}
        \label{fig:vegard_cl}
    \end{minipage}\hfill
    \begin{minipage}[b]{0.5\textwidth}
        \centering
        \includegraphics[width=82mm]{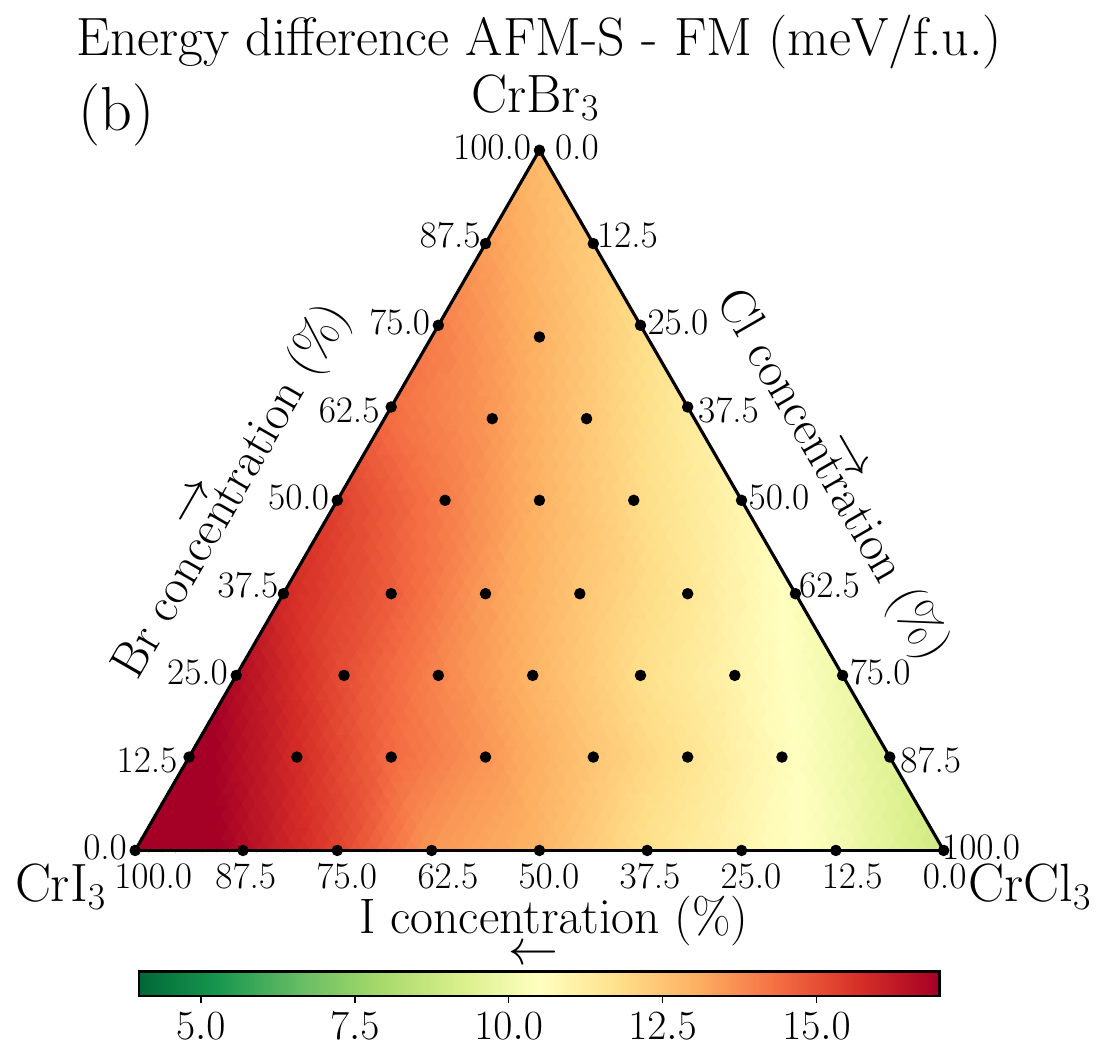}
        \label{fig:vegard_br}
    \end{minipage}

    \vspace{0.4cm}

    \begin{minipage}[b]{0.5\textwidth}
        \centering
        \includegraphics[width=82mm]{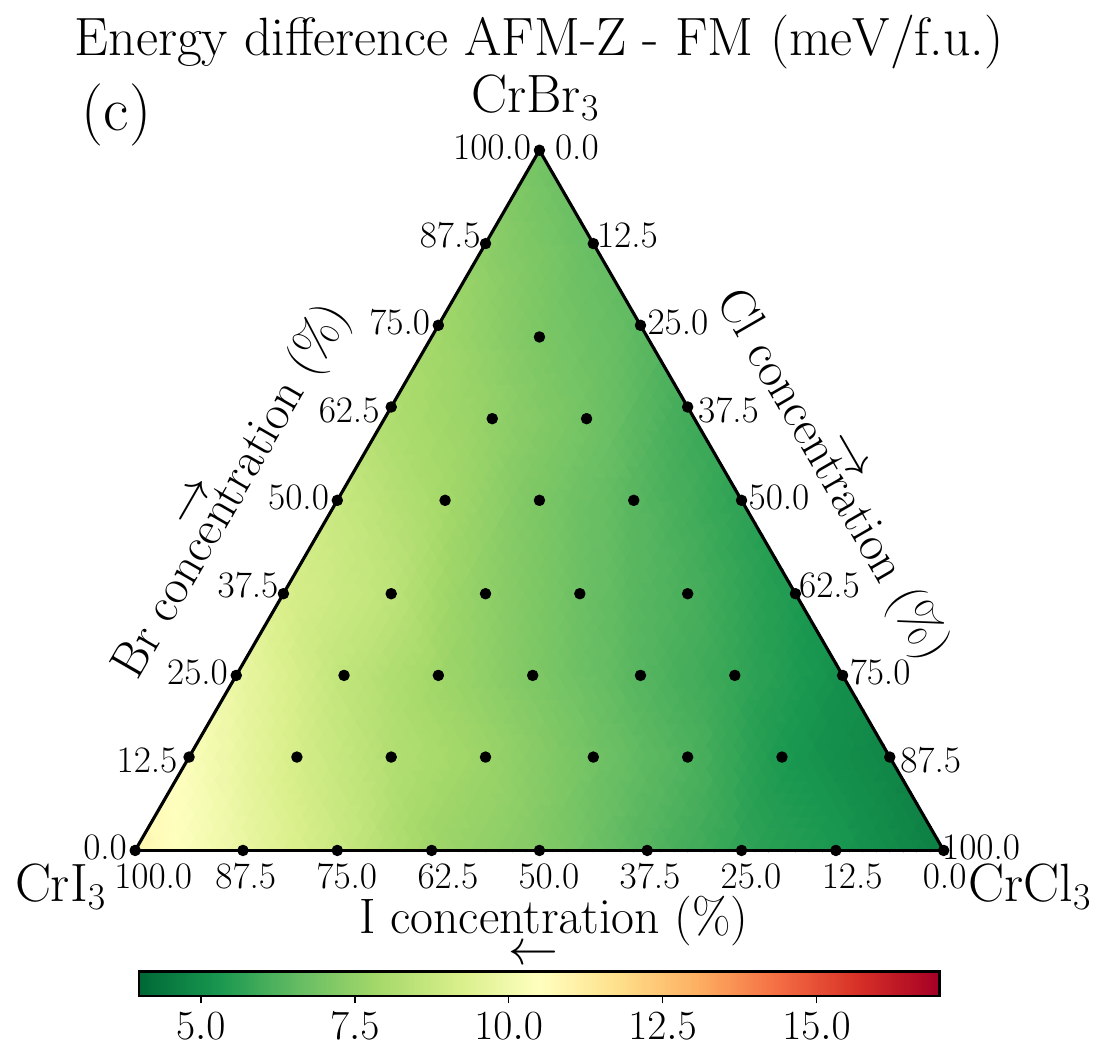}
        \label{fig:vegard_i}
    \end{minipage}\hfill
    \begin{minipage}[b]{0.5\textwidth}
        \centering
        \includegraphics[width=82mm]{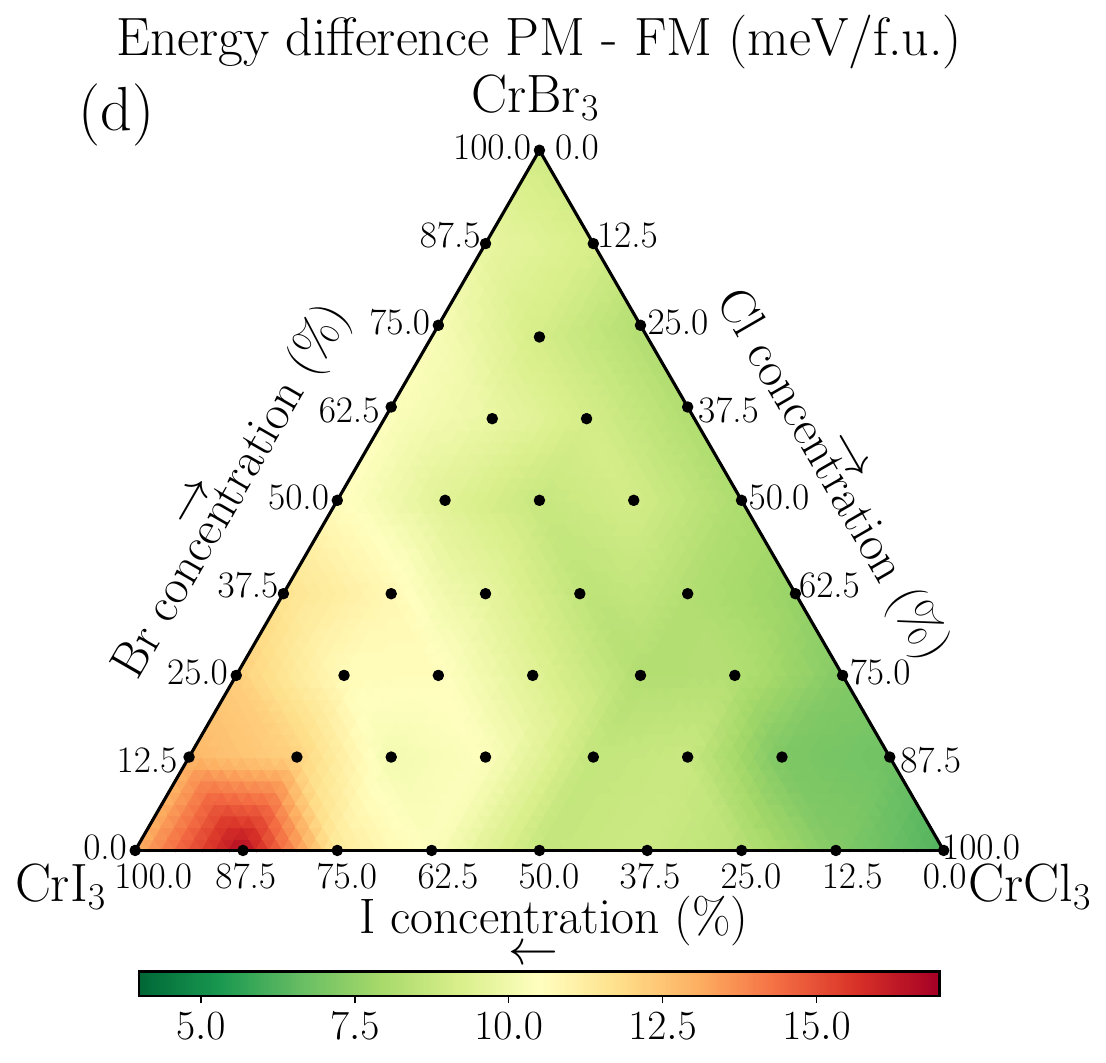}
        \label{fig:vegard_extra}
    \end{minipage}

    \caption{Heat map for the energy difference (in meV/f.u.) between all magnetic phases and FM phase for alloys and pure compounds. The color scale represents the stability trend, with higher values indicating a greater preference for the FM phase. The energy difference varies smoothly across compositions, with the largest values observed near the {CrI}$_{3}$-rich region. Each black point indicates a specific composition; the midpoint of the vertex between CrCl$_{3}$ and CrI$_{3}$ corresponds to Cr(Cl$_{0.50}$Br$_{0.00}$I$_{0.50}$)$_{3}$.}
    \label{energies_differences}
\end{figure}

\begin{figure}[htb]
    \centering
    \begin{minipage}[b]{0.5\textwidth}
        \centering
        \label{fig:bands_li}%
        {\includegraphics[width=82mm]{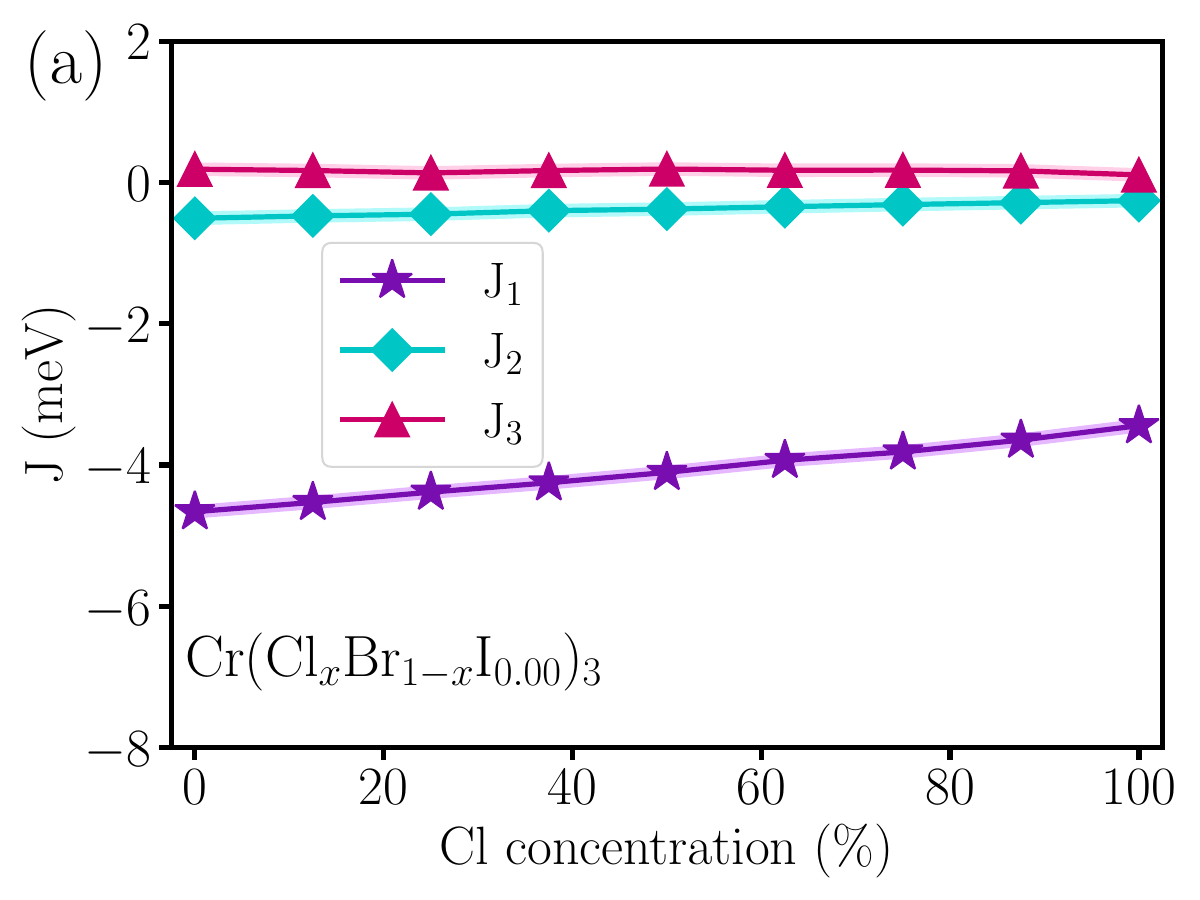}}
    \end{minipage}\hfill
    \begin{minipage}[b]{0.5\textwidth}
        \centering
        \label{fig:bands_k}%
        {\includegraphics[width=82mm]{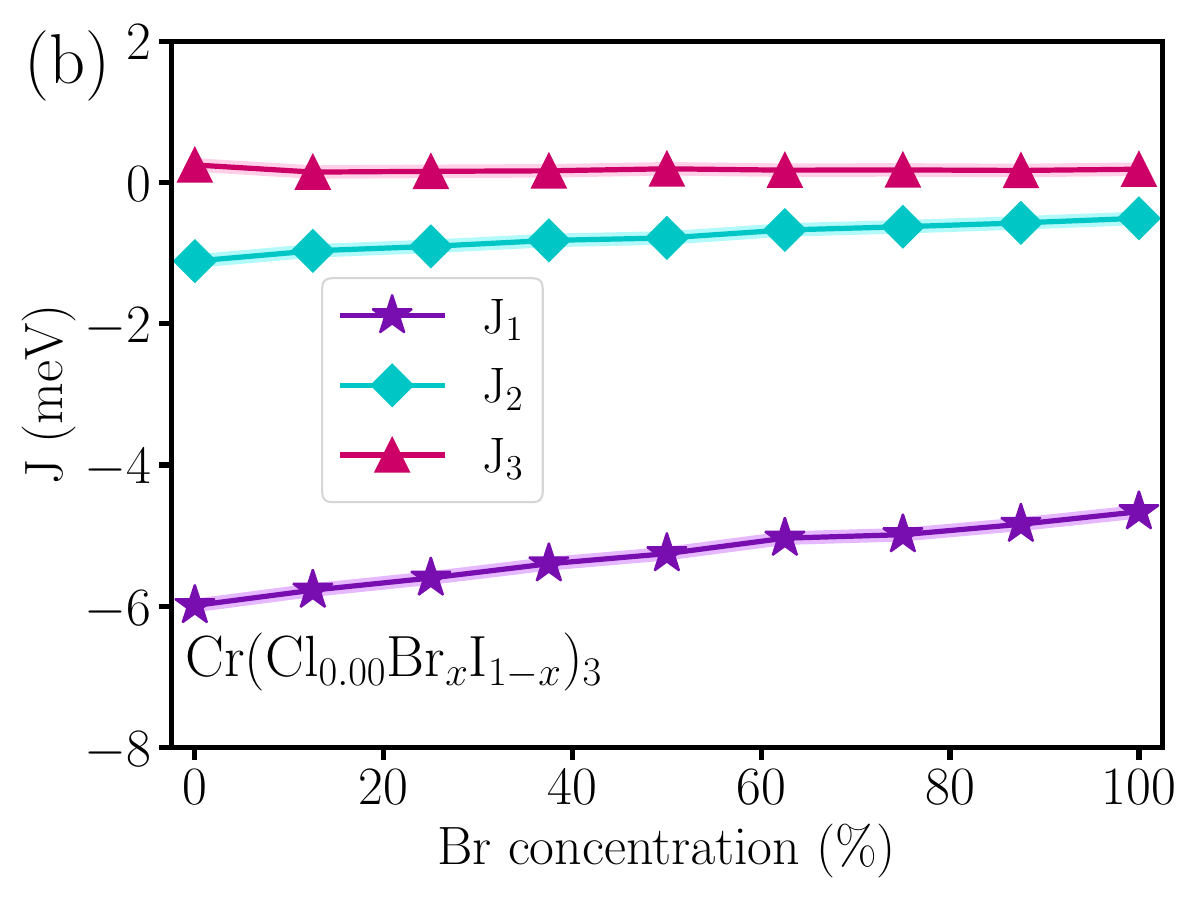}}
    \end{minipage}

    \vspace{0.4cm}

    \begin{minipage}[b]{0.6\textwidth}
        \centering
        \label{fig:bands_na}%
        {\includegraphics[width=82mm]{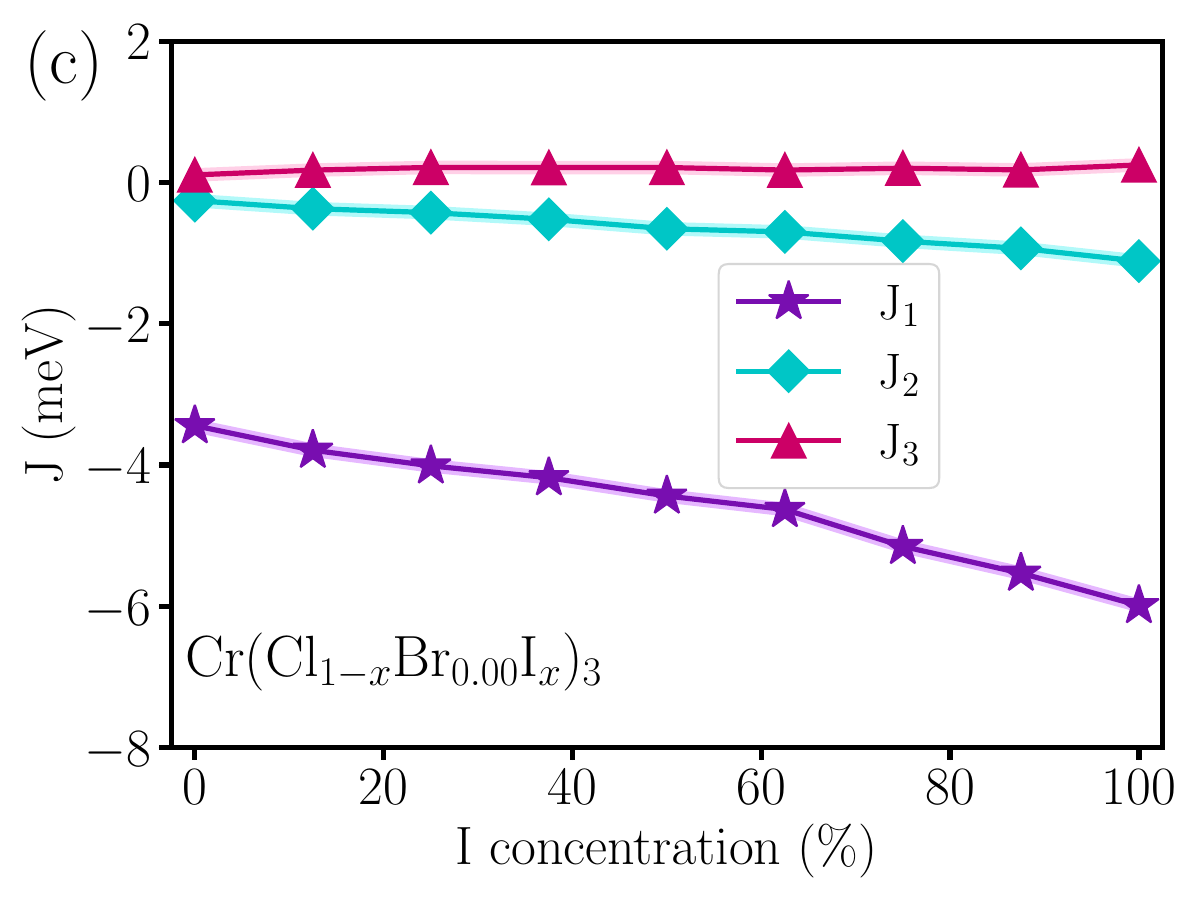}}
    \end{minipage}

    \caption{Magnetic exchange parameters J$_{1}$, J$_{2}$, and J$_{3}$ along the compositional edges connecting the parent compounds. The purple curves with star markers correspond to J$_{1}$, while the light blue curves with diamond markers and the dark pink curves with triangle markers represent J$_{2}$ and J$_{3}$, respectively. The specific compositions are indicated in each panel.}
    \label{Js}
\end{figure}

Figure \ref{energies_differences} shows the heatmaps for the energy difference between (a) AFM-N, (b) AFM-S, (c) AFM-Z, (d) PM phases and the FM ground state. As observed in the heatmaps, the overall trend closely follows that of the pure compounds, with the largest energy differences relative to the FM ground state occurring for the AFM-N configuration, followed by AFM-S, PM, and finally AFM-Z. A clear pattern emerges in which the I-rich region exhibits the highest energy differences, with the maximum value corresponding to the parent compound CrI$_3$, reaching 19.37 meV (see Table I) in Fig. \ref{energies_differences}-(a). Pristine CrI$_3$ is strongly ferromagnetic, and the substitution with Cl and Br alters the magnetic coupling - mainly J$_{1}$ and J$_{2}$ as can be seen in Fig. \ref{Js}-(c) - between Cr atoms, leading to a reduction in the energy differences between AFM and FM phases.

In Figure \ref{energies_differences}-(d), the heatmap for the energy difference between the PM and FM phases shows a more pronounced variation than in Figures \ref{energies_differences} (a)-(c) in the CrI$_3$-rich region. The larger energy difference is located notably in the Cr(Cl$_{0.125}$Br$_{0.00}$I$_{0.875}$)$_3$ alloy, which exhibits an energy difference of 16.70 meV. This enhanced PM–FM energy difference at this specific composition can be attributed to the increased magnetic disorder in the PM phase, which leads to stronger spin fluctuations across the lattice. Moreover, the size mismatch between iodine and chlorine atoms induces local structural distortions, which further affect the exchange interactions between Cr atoms.

\begin{figure}[h!]%
 
       \centering
        \includegraphics[width=100mm]{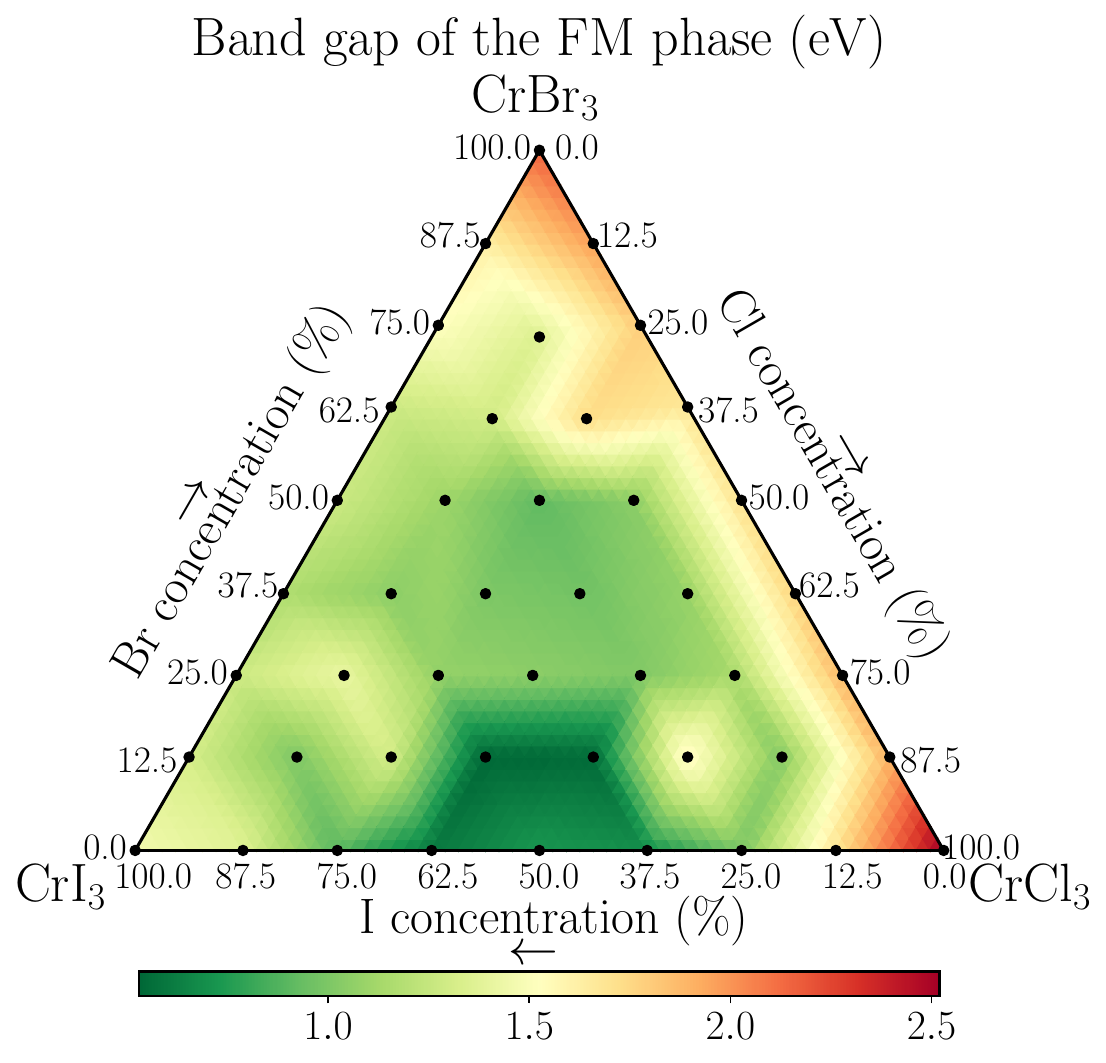}%
         \caption{Heatmap of the band gap (in eV) for the FM phase. Red regions indicate areas with higher band gaps, while green regions indicate areas with lower band gaps.}
        \label{ht_fm}
\end{figure}

Beyond the magnetic energy differences, the composition of the alloys also significantly influences their electronic properties. In particular, the band gap is a key descriptor of the material’s electronic and optical behavior. Figure \ref{ht_fm} shows the heatmap of the band gap for the most stable FM phase in the ternary alloys. The largest band gap values are found along the line connecting the pure compounds based on Cl and Br, which can be attributed to the influence of halogens on the electronic structure. Along this line, the iodine fraction is minimal or absent, suppressing the gap-reducing effect associated with $p$–$d$ coupling.

In Figure \ref{ht_fm}, an uneven variation in the band gap values can be observed at certain points. Interestingly, this behavior is not present in the AFM-Z or PM phases (see the SPM). This suggests that magnetic ordering plays a significant role in shaping the electronic structure of these mixed halide compounds. In the FM phase, the alignment of magnetic moments may enhance the exchange splitting of the Cr $d$ orbitals, making the system more sensitive to the local chemical environment and to variations in halogen composition. In contrast, in the AFM-Z and PM phases, the spin interactions are either compensated or disordered, leading to a more averaged contribution of the halogen orbitals to the band edges and suppressing the localized effects observed in the FM phase.

In mixed halide alloys, halogen orbitals play a crucial role in determining the band gap. As observed in Figure \ref{ht_fm}, a bowing—deviation from linear interpolation—is present in all directions of the compositional space, being more pronounced along the Cl–I edge. Along this direction, the strong mixing between Cl and I leads to a significant bowing of the band gaps. This behavior reflects underlying electronic interactions and highlights the potential for band-edge engineering to tailor the electronic levels for specific applications. In general, iodine’s more delocalized and higher-energy $p$ orbitals tend to reduce the band gap.

\begin{figure}[htb]
    \centering
    \begin{minipage}[b]{0.5\textwidth}
        \centering
        \label{fig:bands_li}%
        {\includegraphics[width=82mm]{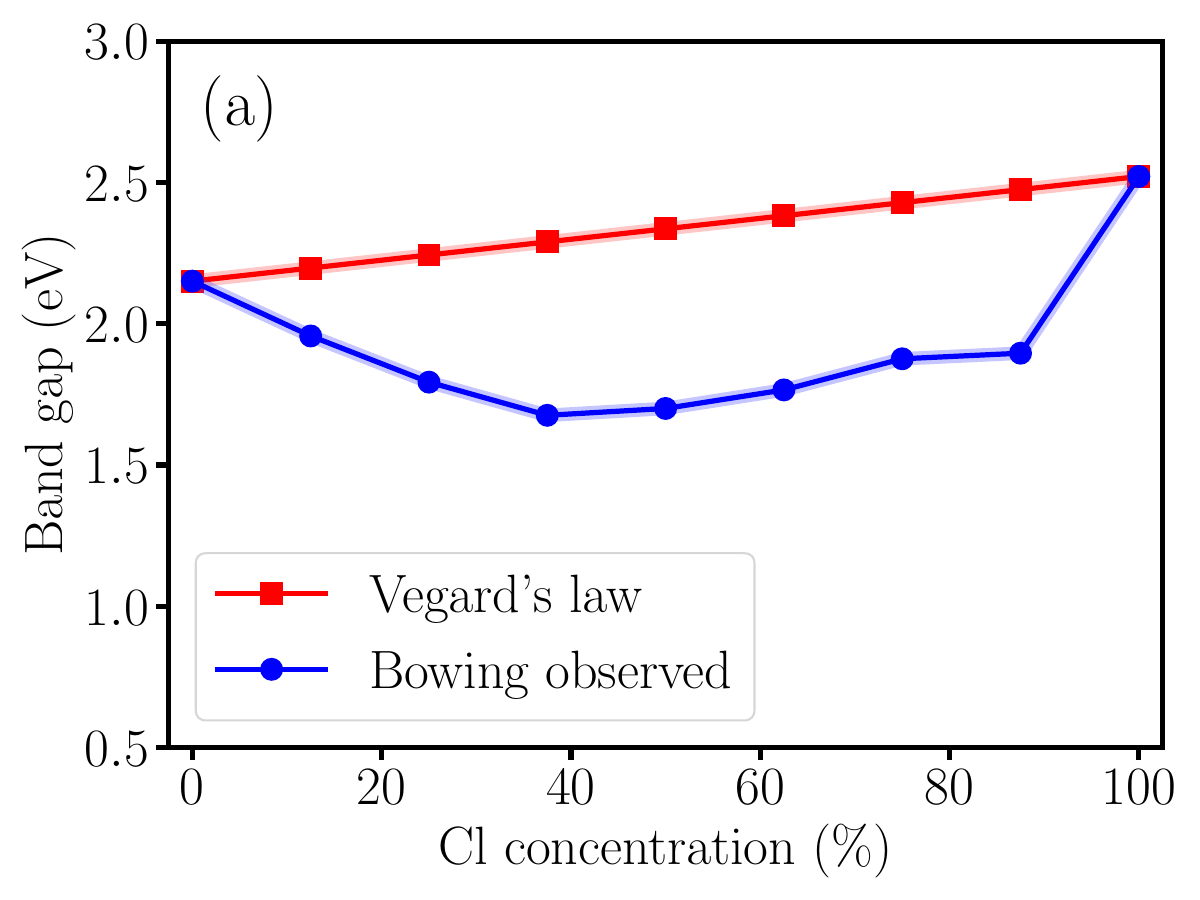}}
    \end{minipage}\hfill
    \begin{minipage}[b]{0.5\textwidth}
        \centering
        \label{fig:bands_k}%
        {\includegraphics[width=82mm]{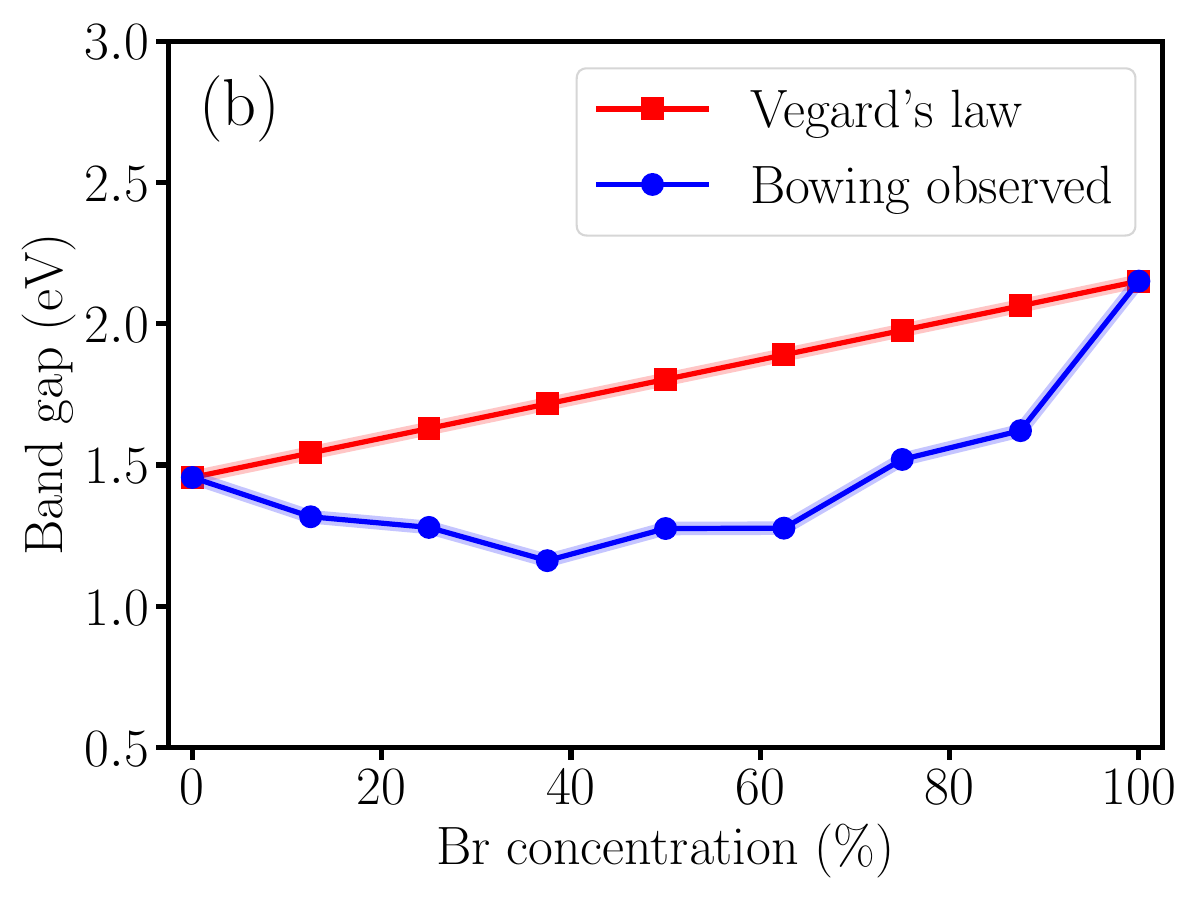}}
    \end{minipage}

    \vspace{0.4cm}

    \begin{minipage}[b]{0.6\textwidth}
        \centering
        \label{fig:bands_na}%
        {\includegraphics[width=82mm]{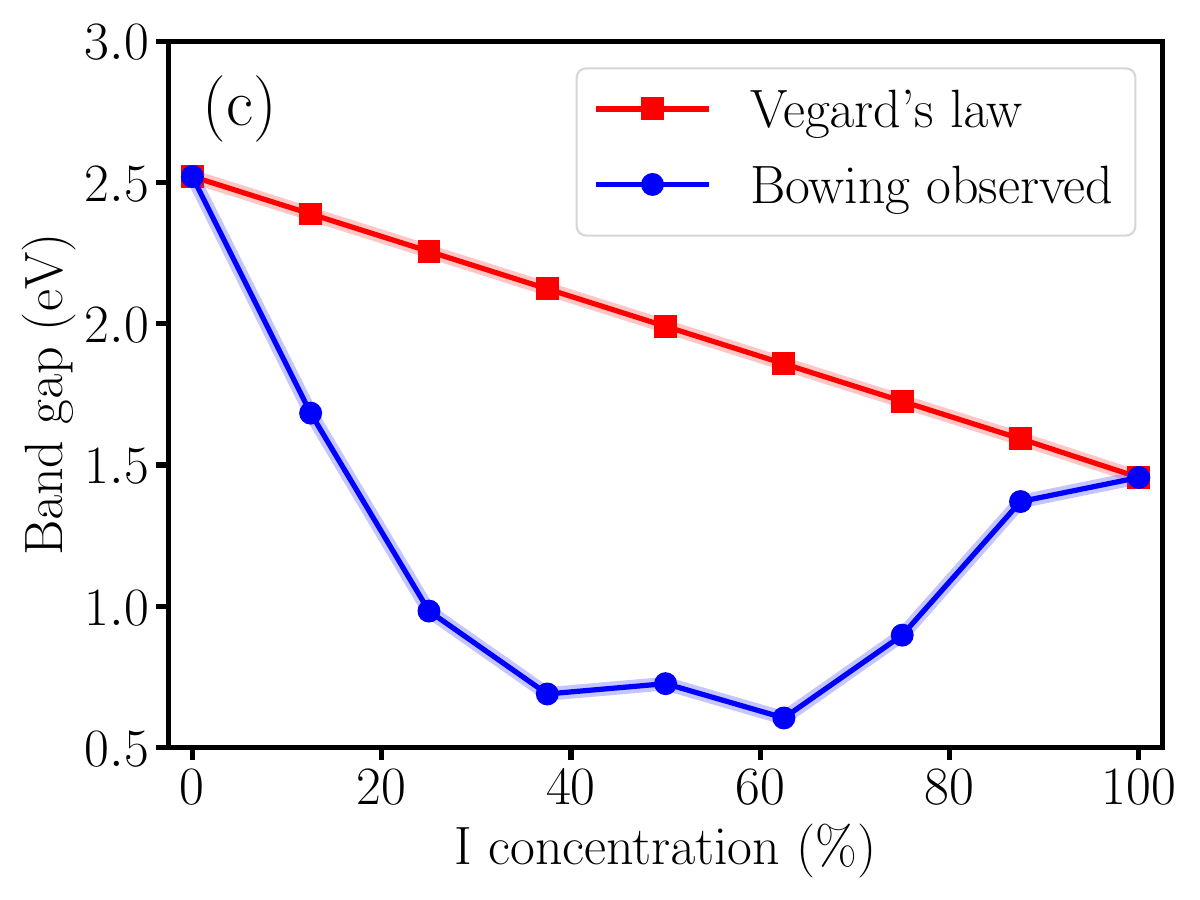}}
    \end{minipage}

    \caption{Quantification of the observed bowing at the Br–Cl (a), I–Br (b), and Cl–I (c) vertices. The blue line represents Vegard’s law, while the red line shows the observed bowing. The Vegard’s law trends reflect the gaps of the pure compounds: CrCl$_{3}$ (2.52 eV), CrBr$_{3}$ (2.14 eV), and CrI$_{3}$ (1.46 eV). Thus, increasing the fraction from Br to Cl or from I to Br enlarges the band gap, while increasing from Cl to I reduces it.}
    \label{bow}
\end{figure}

To gain deeper insight into how the electronic structure evolves across compositions, the band gaps were analyzed for the three binary alloy systems. As shown in Figure \ref{bow}, the evolution of the band gap exhibits a clear non-linear behavior in all directions of the ternary diagram, with the strongest deviation occurring along the Cl–I edge. This nonlinearity underscores the complex interplay between local atomic environments and electronic hybridization in mixed halide systems.

The maximum deviation is found for the Cr(Cl$_{x}$Br$_{0.00}$I$_{1-x}$)$_{3}$ system, with a bowing value of 1.43 eV, followed by 0.63 eV for Cr(Cl$_{x}$Br$_{1-x}$I$_{0.00}$)$_{3}$ and 0.55 eV for Cr(Cl$_{0.00}$Br$_{x}$I$_{1-x}$)$_{3}$. The maximum bowing, in the I-Cl line, occurs before the equimolar composition, indicating an asymmetric substitution effect, likely linked to the differences in electronegativity and ionic radius among the halogens. This behavior is further influenced by structural distortions induced by halogen substitution. As previously mentioned, iodine is significantly larger and less electronegative than chlorine, so gradually replacing Cl with I introduces local distortions in the crystal lattice. For instance, in the relaxed structures of the parent compounds, the Cr–Cr distance is 3.46 Å in CrCl$_3$ and 4.02 Å in CrI$_3$, with Cr–Cl–Cr and Cr–I–Cr angles of 94.96° and 94.59°, respectively. In the Cr(Cl$_{0.50}$Br$_{0.00}$I$_{0.50}$)$_3$ alloy, these parameters differ significantly (Cr–Cr = 3.82 Å, $\phi$ = 103.91°, $\theta$ = 87.12°), introducing lattice strain. Such distortions modify the local electronic environment, influencing exchange interactions and contributing to the observed bowing of the band gap.

\begin{figure}[h!]%
 
       \centering
        \includegraphics[width=100mm]{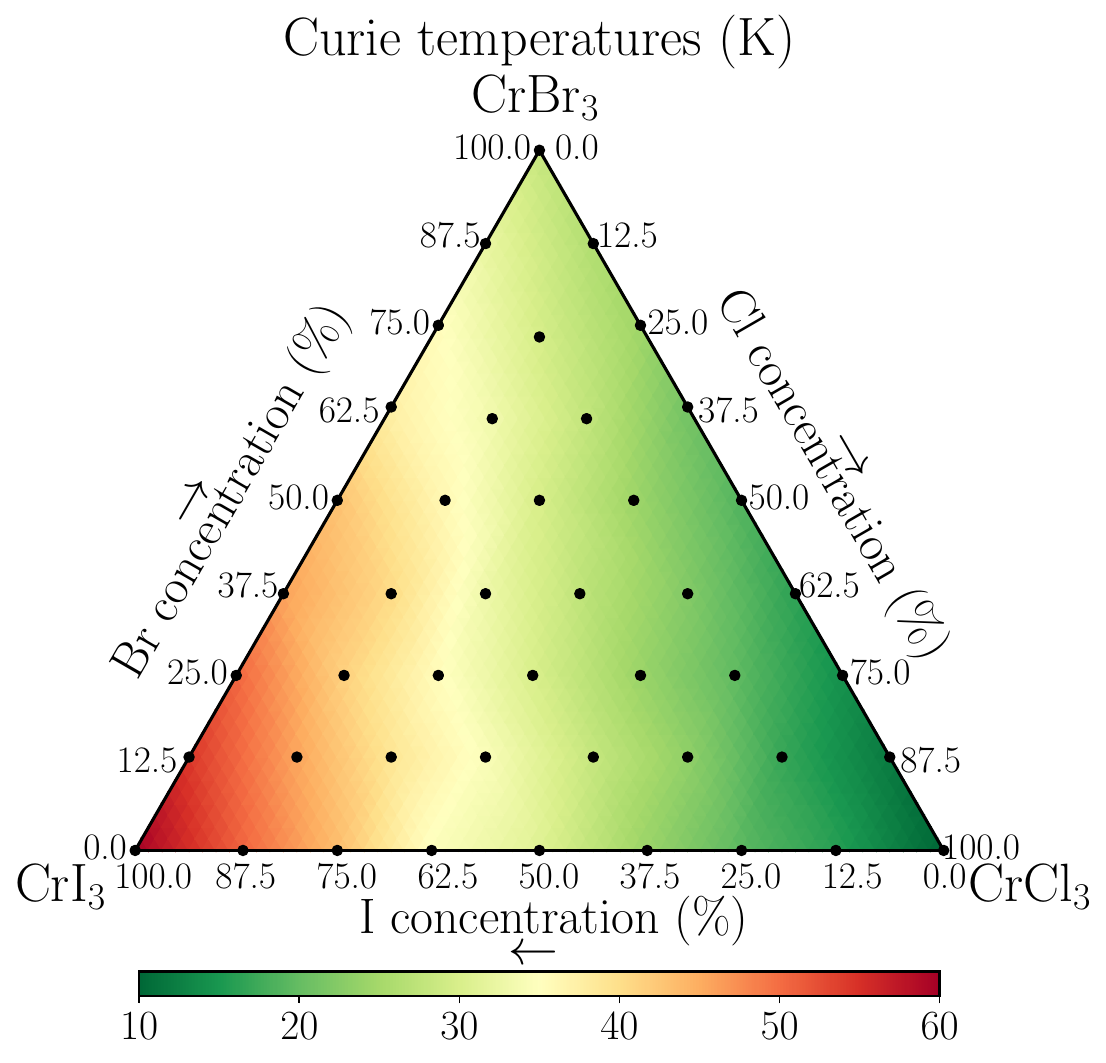}%
         \caption{Curie temperatures across the compositional space of CrX$_3$ alloys. A smooth variation is observed throughout the phase diagram, with the lowest values approaching the CrCl$_3$ vertex and the highest values near the CrI$_3$ vertex, reflecting the compositional dependence of the magnetic interactions.}
        \label{TCs_alloys}
\end{figure}

As previously mentioned, the bulk phases of the parent compounds exhibit relatively low Curie temperatures, indicating that their alloys may also become paramagnetic at room temperature. Figure \ref{TCs_alloys} presents the calculated Curie temperatures for all alloy compositions. As shown, T$_{c}$ varies smoothly across the composition space, with the highest values in the I-rich region, intermediate values near the center of the heatmap, and the lowest values in the Cl-rich region. This trend is consistent with experimental observations \cite{Tartaglia2020}; however, our results provide additional insight into the Cl–I compositional edge, which could not be observed experimentally\cite{Tartaglia2020}. The maximum and minimum Curie temperatures obtained are 51.42 K and 22.22 K, corresponding to the CrI$_3$ and CrCl$_3$ vertices, respectively, in good agreement with the literature\cite{Xue2022}. These findings demonstrate that alloy composition can be effectively used as a tuning parameter for the Curie temperature.

To gain deeper insight into the Curie temperatures, the magnetic exchange parameters were evaluated. Figure \ref{Js} presents the exchange interactions along the (a) Br–Cl, (b) I–Br, and (c) Cl–I compositional edges. As shown, the nearest-neighbor interaction J$_{1}$ exhibits the most significant variation, followed by the next-nearest-neighbor interaction J$_{2}$, while J$_{3}$ interaction remains nearly constant and close to zero. A clear increasing trend in both J$_{1}$ and J$_{2}$ is observed along the Br–Cl and I–Br edges, whereas a decreasing occurs along the Cl–I edge, consistent with the higher Curie temperatures in this region. Overall, the exchange parameters vary approximately linearly across these binary edge alloys, in agreement with the smooth behavior observed in the Curie temperature heatmap. These results indicate that the first- and second-neighbor interactions dominate the magnetic behavior and are primarily responsible for the gradual tuning of T$_{C}$ across the alloy compositions.

In addition to electronic properties and Curie temperatures, assessing the thermodynamic stability of the Cr(Cl$_x$Br$_y$I$_{1-x-y}$)$_3$ alloys provides insight into their likelihood of formation and persistence. This stability is estimated through the Gibbs free energy, calculated from the mixing enthalpy and the configurational entropy contribution. Here, only the configurational term is considered, neglecting vibrational, translational, and rotational contributions. Despite this simplification, the approach captures the essential trends in the energetic favorability of the alloys across compositions. The Gibbs free energy is approximated as\cite{Porter1992}:

\begin{center}
\begin{equation}
\displaystyle \Delta G =\Delta H_{mix} - T\Delta S_{mix}
\end{equation}
\end{center}
where $\Delta H_{mix}$ is the mixing enthalpy, $T$ is the temperature and $\Delta S_{mix}$ is the mixing entropy. 

While the mixing enthalpy determines the energetic favorability of alloy formation, the configurational entropy accounts for the disorder associated with different atomic arrangements. Together, these contributions allow the assessment of thermodynamic stability and phase behavior through the Gibbs free energy. For the alloys, $\Delta G$ reflects the balance between enthalpic and entropic effects: even positive mixing enthalpies can lead to stable mixed phases at sufficiently high temperatures due to the entropic contribution. The mixing enthalpy is calculated as:

\begin{center}
\begin{equation}
\displaystyle \Delta H_{mix} =E_{alloy} -\eta ( \alpha E_{Cl} +\beta E _{Br} +\gamma E_{I})
\end{equation}
\end{center}
where $E_{alloy}$, $E_{Cl}$, $E_{Br}$, and $E_{I}$ are the energies of the alloy and the pure compounds {CrX}$_{3}$. The parameters $\alpha$, $\beta$, and $\gamma$ represent the respective concentrations of the halides in the alloys, and $\eta$ is a normalization term accounting for the supercell size of the alloy, as different supercell (volumes) sizes were used for the different magnetic phases (see the SPM). The mixing entropy can be calculated using the equation\cite{Mohammadzadeh2023}:

\begin{center}
\begin{equation}
\displaystyle \Delta S_{mix} =-R \sum _{i} x_{i}\ln x_{i}
\end{equation}
\end{center}
where R is the ideal gas constant and $x_{i}$ is the concentration of each halogen is the concentration of each halogen in the alloy - in the alloy Cr(Cl$_{0.875}$Br$_{0.125}$I$_{0.00}$)$_{3}$ $x_{i}$ has the values: 87.5, 12.5 and 0.0 for instance.

\begin{figure}[htb]
    \centering

    \begin{minipage}[b]{0.5\textwidth}
        \centering
        \includegraphics[width=82mm]{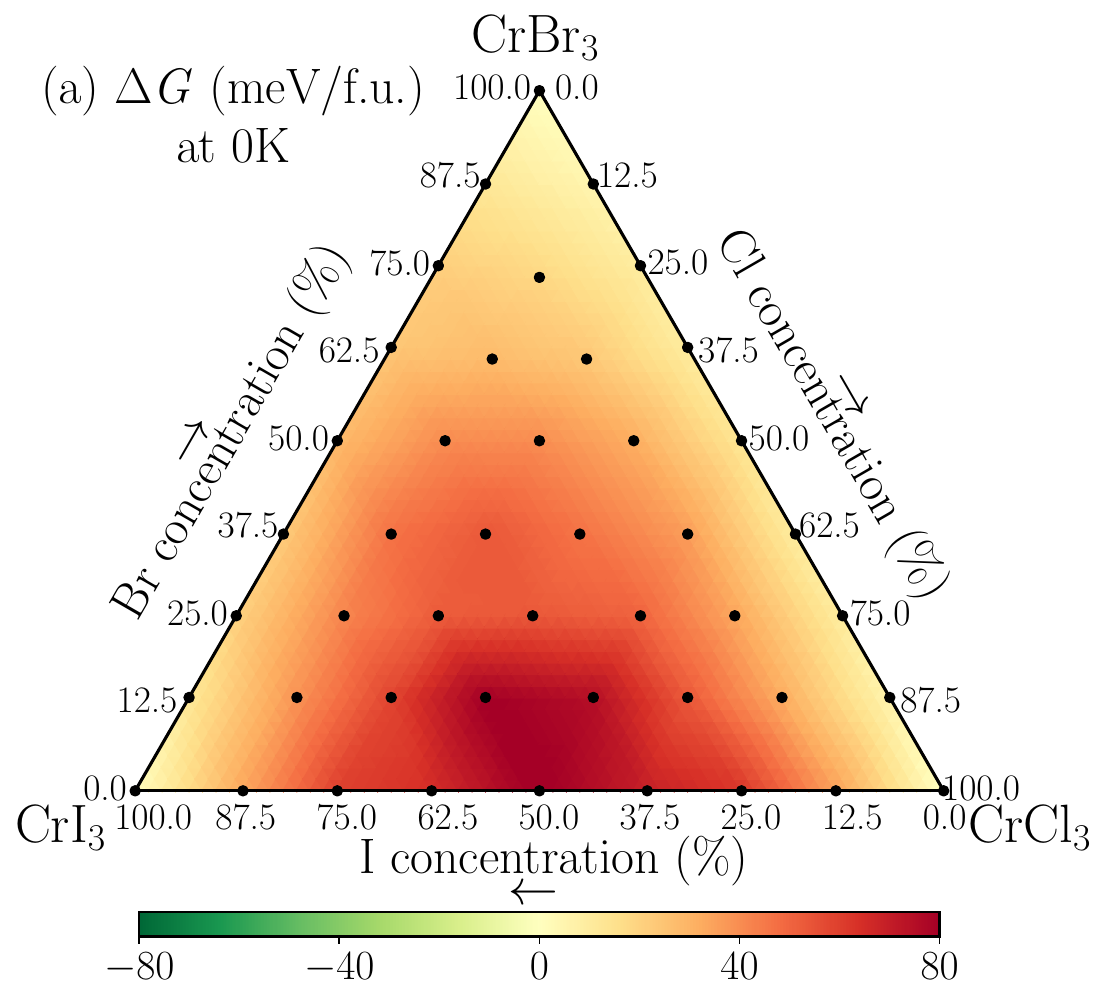}
        \label{fig:vegard_cl}
    \end{minipage}\hfill
    \begin{minipage}[b]{0.5\textwidth}
        \centering
        \includegraphics[width=82mm]{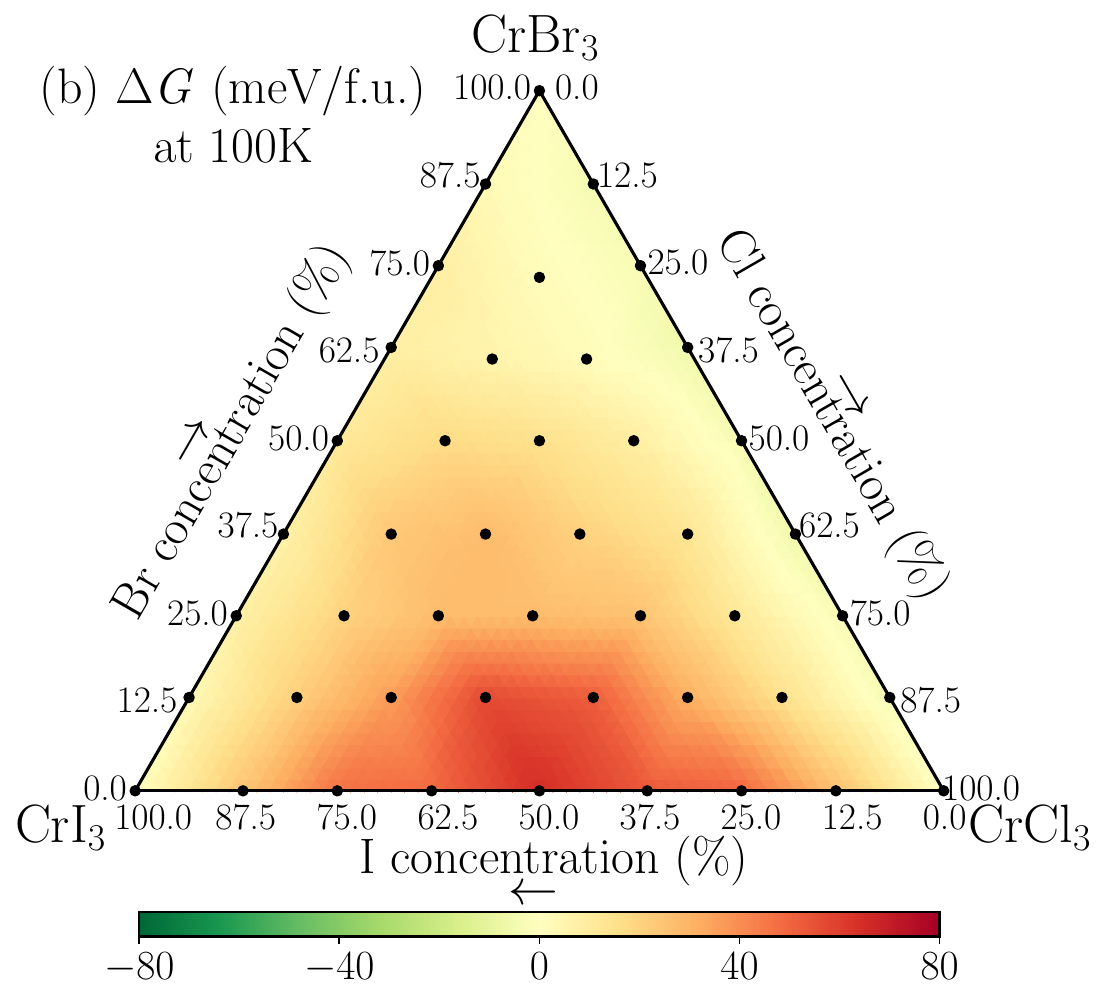}
        \label{fig:vegard_br}
    \end{minipage}

    \vspace{0.4cm}

    \begin{minipage}[b]{0.5\textwidth}
        \centering
        \includegraphics[width=82mm]{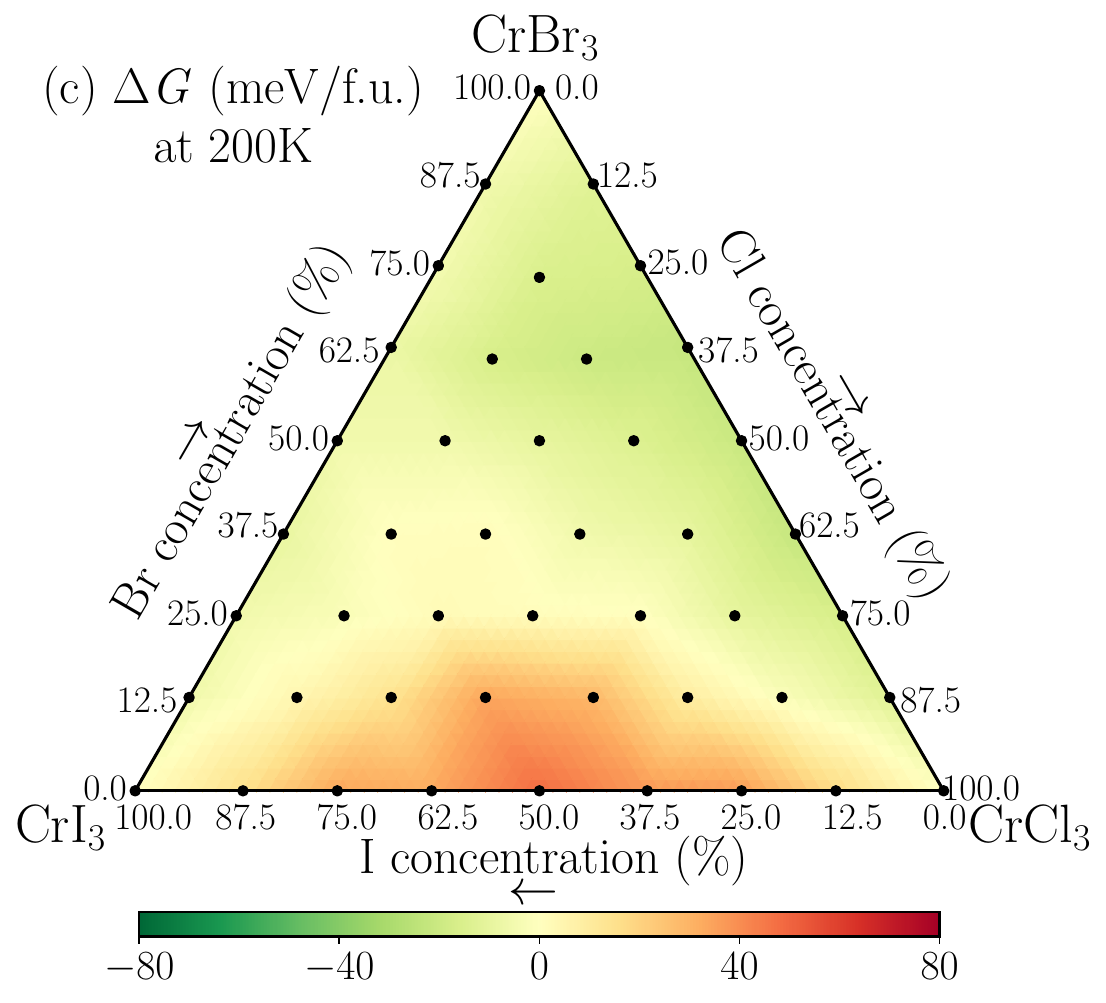}
        \label{fig:vegard_i}
    \end{minipage}\hfill
    \begin{minipage}[b]{0.5\textwidth}
        \centering
        \includegraphics[width=82mm]{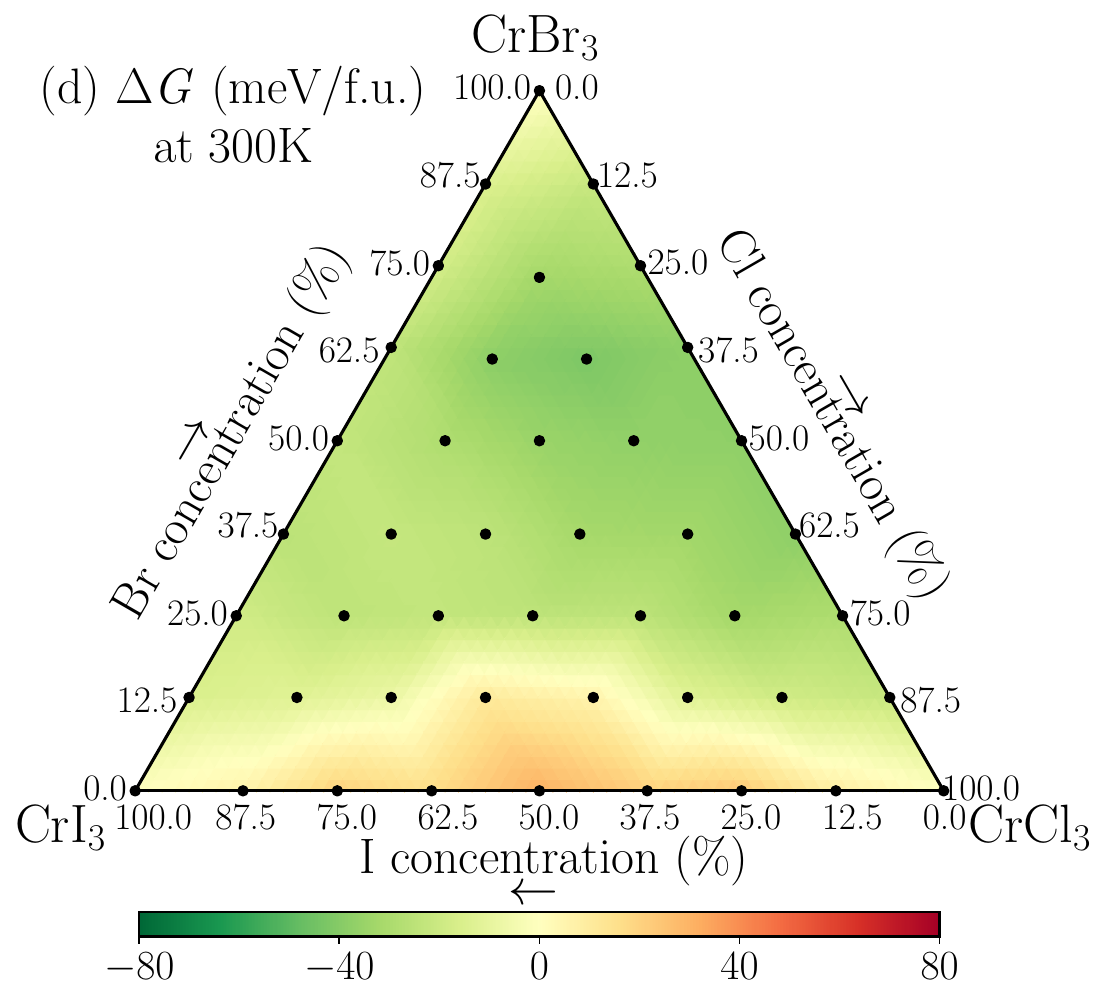}
        \label{fig:vegard_extra}
    \end{minipage}

    \caption{Gibbs free energy in meV/f.u. obtained for different temperatures. Green regions highlight compositions with a thermodynamically favored structure, while red regions highlight higher energies.}
    \label{egsc}
\end{figure}

The Gibbs free energy of the Cr(Cl$_x$Br$_y$I$_{1-x-y}$)$_3$ alloys was first calculated at 0 K to isolate the effect of the mixing enthalpy, neglecting any entropic contributions. Subsequently, the free energy was evaluated at 100, 200, and 300 K, including the configurational entropy contribution. The resulting temperature-dependent trends in $\Delta G$ across compositions are presented in Figure \ref{egsc}.

Figure \ref{egsc} shows the heatmaps of the Gibbs free energy at different temperatures. As temperature increases, $\Delta \text{G}$ becomes negative over a significant portion of the compositional space, indicating that the configurational entropy progressively compensates for the positive mixing enthalpy, making alloy formation thermodynamically favorable at sufficiently high temperatures. This implies that some compositions unstable at 0 K can stabilize under synthesis or annealing conditions due to entropy-driven lowering of the free energy.

In Figure \ref{egsc}-(a), where $T=0$ K and $\Delta G$ reduces to the mixing enthalpy $\Delta H_{mix}$, the highest values are found along the CrCl$_3$–CrI$_3$ edge. All calculated $\Delta H_{mix}$ values are positive, with a maximum of 81 meV/f.u. for Cr(Cl$_{0.50}$Br$_{0.00}$I$_{0.50}$)$_3$, suggesting reasonable thermodynamic stability, particularly when compared to typical unstable systems that exhibit much larger mixing enthalpies ($>$0.2–0.3 eV/f.u.) but are experimentally observed as single-phase alloys\cite{Manzoor2018,Zaddach2013}. As can also be seen in Figure \ref{egsc}-(a), the lowest $\Delta G$ values are located along the CrCl$_{3}$–CrBr$_{3}$ edge, indicating that alloying in compositions such as Cr(Cl$_{0.50}$Br$_{0.50}$I$_{0.00}$)$_3$ strongly favors the thermodynamic stabilization of these structures. This behavior may help explain why such alloys have been experimentally realized\cite{Tartaglia2020}.

Figure \ref{egsc}-(d) shows that compositions with high configurational disorder, particularly near equiatomic ratios, gain the largest stabilizing contribution from $\Delta S_{mix}$. Regions with high configurational disorder stabilize at lower temperatures, while less mixed regions require higher temperatures to achieve $\Delta G<0$. As can also be observed in this figure, even at room temperature, the CrCl$_{3}$–CrI$_{3}$ edge still exhibits positive or near-zero $\Delta G<0$ values, indicating that these alloys are thermodynamically unfavorable or only marginally stable. This behavior helps explain why compositions such as Cr(Cl$_{0.50}$Br$_{0.00}$I$_{0.50}$)$_3$ have not been experimentally realized\cite{Tartaglia2020}. Therefore, the absence of these alloys is not solely due to the atomic radius mismatch between Cl and I, but also originates from this fundamental thermodynamic constraint.

\newpage

\section{Conclusions \label{Conclusions}}

In summary, we have theoretically studied the {CrX}$_{3}$ compounds, where X = Cl, Br, and I, as well as their ternary alloys, using DFT calculations. We found that, for the pure compounds, the ground state is intrinsically FM, while the AFM-Z and PM phases are close in energy. The band gaps for the pure compounds vary by at most 0.16 eV across different magnetic phases and the character of the VBM changes depending on the halide: for CrCl$_{3}$, the Cr d orbitals dominate the VBM; for CrBr$_{3}$, there is significant hybridization between Br p and Cr d states; and in CrI$_{3}$, the I p orbitals overwhelmingly contribute to the VBM. Meanwhile, at the CBM, the Cr d orbitals remain the primary contributors across all three materials. We explain the different band gaps between halogens through $p-d$ coupling, so that as one moves from Cl to Br and then to I, the p-orbital energy of the halogens increases and the orbital itself becomes more delocalized. This higher energy and greater diffuseness enhance the overlap with the Cr d orbitals, effectively
pushing the valence-band maximum upward in energy. The largest energy difference between the FM ground state and the PM phase is only 13.01 meV per formula unit. Such a small difference is consistent with the low Curie temperatures of these monolayers: as thermal energy increases, one can anticipate sequential transitions from the FM ground state to the AFM-Z configuration and, eventually, into the high-temperature PM regime. 

For the alloys, the FM state remains the ground state, with compositions containing less iodine exhibiting energy values very close to the AFM-Z phase. The largest band gap values are found along the edge
connecting the parent compounds based on Cl and Br. Along the edge connecting Cl and Br, the iodine fraction is minimal or absent, eliminating the gap-reducing effect provided by $p-d$ coupling. A bowing is observed along the edge connecting the original compounds Cl and I, that is, the alloy Cr(Cl$_{x}$Br$_{0.00}$I$_{1-x}$)$_{3}$. The largest deviation occurs when the iodine concentration is 37.5\% and the chlorine concentration is 62.5\%, providing a difference of 1.43 eV from linear interpolation between the systems (Vegard's law). The Curie temperatures were also determined, showing a smooth variation across compositions, consistent with the nearly linear behavior of the magnetic exchange parameters along the edges connecting the parent compounds. The Gibbs free energy was calculated and our results show that at room temperature most of the alloys, with the exception of those located at the Cl-I edge are stable, indicating reasonably good thermodynamic stability. his effect is particularly pronounced for compositions such as Cr(Cl$_{0.50}$Br$_{0.50}$I$_{0.00}$)$_3$, which exhibit very low Gibbs free energy values even at $T$ = 0 K, strongly favoring the thermodynamic stabilization of these structures. This work provides fundamental insights into the electronic, magnetic, and structural behavior of two-dimensional magnetic materials CrX$_{3}$ and their ternary alloys, underscoring their promise for next-generation spintronic and optoelectronic applications.
We also suggest alloying as a viable strategy to increase the thermodynamic stability of these compounds, substantially increasing the potential for practical applications.

\begin{acknowledgments}
We gratefully acknowledge financial support from CNPq under Grant No. 141697/2023-7 and FAPESP under Grant No. 2023/03493-0 and 2023/09820-2 and 2024/21550-3. The authors acknowledge the computational resources provided by the Laboratório Nacional de Computação Científica (LNCC) and CENAPAD/Campinas.
\end{acknowledgments}

\newpage
\bibliography{ref}

\end{document}